# Roll-to-roll tomographic volumetric additive manufacturing

Joseph Toombs[a], Chi Chung Li[a], Nour Akiki[a], Yaxuan Sun[a], Hayden Taylor[a]

[a]Dept. of Mechanical Engineering, University of California, Berkeley, CA USA 94720

**Abstract**

Additive manufacturing (AM) has revolutionized the fabrication of devices with precisely controlled optical, fluidic, mechanical, and filtering properties, offering greater design freedom than conventional manufacturing methods. Tomographic volumetric additive manufacturing (VAM) has many advantages compared to other AM methods including smooth layer-less surfaces, support-free and shear force-free printing, material versatility, and speed of production which are translatable to the microscale and evident in printed microfluidic devices and micro-optical components. However, as the patterning scale is reduced, the depth of field of the optical projection system shrinks much more rapidly and does so roughly with the square of the patterning scale. Consequently, the build volume is substantially reduced as the numerical aperture of the system is increased. Additionally, microscale tomographic VAM is currently limited to batch production, i.e., the photoresist container must be exchanged after the exposure phase is completed. In this work, we introduce roll-to-roll (R2R) tomographic VAM in which these limitations are addressed by "unwrapping" the precursor material into a film enabling continuous production of microstructures with theoretically unlimited length. We elaborate the design of a focus-multiplexed projection optical system that can scan the projection focal plane axially in sync with the refresh cycle of a digital micromirror device. We describe the process of iteratively optimizing and segmenting sinograms to produce long aperiodic microstructures with the focus-tunable optical system. Furthermore, we formulate a thermally reversible organogel photoresist which is deposited onto the substrate in films several millimeters in thickness. Finally, we demonstrate printing of complex lattice structures with length more than 60 cm and minimum feature size of 200 µm with the R2R tomographic VAM system.

## 1. Introduction

Applications demanding precise control over light propagation, fluid transport, mechanical properties, and particle retention have been revolutionized by the introduction of 3D printing technologies, which grant greater design flexibility than traditional manufacturing [1–4]. Examples of these applications include micro-optics, microfluidic devices, mechanical metamaterials, and membrane filters. Layer-by-layer, light-based 3D printing methods, i.e., stereolithography and digital light processing (DLP) 3D printing, are often used to demonstrate proof-of-concept for such applications. However, they are limited in terms of industrial application because of limited build volume and manual post-processing steps including sacrificial support removal. To realize fully the potential of microarchitected films, scalable manufacturing processes without these limitations are essential.

Layer-by-layer, light-based 3D printing has been adapted to continuous processing by companies including Sprybuild and OPT Industries. Printing onto tilted continuous film substrates is possible by adopting (from DLP printing) a top-down approach where the layer is printed at an air-photoresist interface onto a film which gradually descends into a vat, or a bottom-up approach where the light pattern is projected through a window––possibly oxygen-permeable or heated to prevent adhesion to polymerized material—onto a film which gradually ascends out of a vat [5,6]. However, like DLP printing, these techniques can be limited by large suction forces when a window is used or slow speeds when incremental film descent-ascent steps are needed to recoat the printed construct with fresh layers of resin.

Continuous liquid interface production (CLIP) 3D printing was recently demonstrated on a roll-to-roll (R2R) system for high-throughput fabrication of shape-specific particles [7]. In this implementation, a portion of the flexible substrate served as a printing platform such that particles could be printed while the film was paused and displaced vertically over a light projection system. Although this method enabled automated high-throughput production, it was limited to fabrication of discontinuous batches of particles which fit within the field of view of a single light projection due to the stop-start motion of the film.

Volumetric additive manufacturing (VAM) processes—where material is selectively transformed within a volume—include computed axial lithography (CAL), xolography, and two-photon polymerization (2PP). VAM processes are advantageous for continuous printing because photoresist recoating steps and suction forces are eliminated, enabling high printing speed of delicate structures. Additionally, VAM processes allow for the embedding of premanufactured functional objects in the precursor material, and they offer high geometric flexibility because sacrificial supports are not required for many structures [8–11].

CAL is a tomographic VAM method which enables volume-at-once 3D printing of whole objects with high geometric flexibility [9,10]. In our previous work, we have printed objects with minimum feature size of 20 µm in a free-radical polymerization photoresist with a refined system called micro-CAL [12]. To achieve higher printing resolution, the numerical aperture (NA) of the projection optical system was increased. Since the depth of focus has dependence on the inverse square of NA, increasing NA resulted in substantially reduced build diameter for CAL; the maximum usable diameter is restricted by the tolerable defocus of the projection optical system (even if the lateral size of the image is larger than this value). Therefore, objects with large cross-section and small feature size are difficult to produce with micro-CAL.

Xolography, a linear translational volumetric 3D printing process, is made possible by the use of photochromic sensitization of a photoinitiator [11]. A light sheet of one color sensitizes a plane of photoresist and an orthogonally propagating cross-sectional image of an object of another color initiates polymerization. In the seminal work, the photoresist was translated relative to the stationary light sheet. In a second implementation, the photoresist was instead pumped through a specially designed flow cell to extend the build volume along the direction of flow to infinity, theoretically [13]. In this configuration, objects with minimum lateral feature size of 10 µm were printed at speeds up to 3.5 mm $min^{-1}$. Although xolography is a flexible process, it requires special types of photoinitiators, a dual-wavelength projection system, and dual-wavelength transmission of the photoresist.

At a smaller scale, for 2PP, typically either the photoresist or objective lens is translated by a piezo stage. Piezo elements have limited range and, thus, to extend printing height a larger coarse stage must be combined with or take the place of the piezo stage. To address this limitation, 2PP has been adapted to continuous fabrication with clever microfluidic delivery of fresh photoresist. Long microtubule scaffolds (100:1 aspect ratio) for vascular tissue engineering were printed at speeds of 12.5 μm $s^{-1}$ in a vertical flow microfluidic chip [14]. Furthermore, 3D colloidal microparticles were printed in a horizontal flow channel at rates of 30 particles per second [15]. To achieve volumetric patterning these continuous printing architectures require femtosecond lasers, which are expensive and require amplification if optically expanded to large-scale cross-sections for advanced, high-throughput techniques.

Several VAM methods have demonstrated high printing resolution and precise control of part functionality; however, it is challenging to scale these methods in such a way that this local resolution is maintained in objects with large overall dimensions due to specialized optical systems and/or requirements for high photoresist optical transmission at the actinic wavelength [16]. In this work we introduce R2R CAL, which combines the benefits of continuous R2R processing with the light-based tomographic printing approach of conventional CAL to address the scalability challenges and to create 3D structures without supports within a layer of photoresist on a flexible substrate. R2R CAL exploits free-radical polymerization inhibition and therefore does not require a femtosecond laser source or a specialized dual-color photoinitiator to achieve continuous volumetric photopolymerization. The motion of the photoresist around a conveyor roller grants access to azimuthal projections relative to the photoresist coordinates. We introduce a varifocal optical system and depth-multiplexed projection method to create fine scale patterns on different image planes which conform to the shape of this conveyor roller. To support continuous printing of long structures, we explicitly model optical dose accumulation process algebraically to facilitate computationally efficient projection optimization. Finally, we demonstrate printing of lattice structures with sub-millimeter minimum feature size and the largest length dimension by a tomographic patterning method to date.

## 2. Results and Discussion

*R2R CAL concept*

In the conventional CAL configuration, the optical system is typically focused on the rotation axis and structures are also printed in the region close to the rotation axis. In fact, structures can be printed with the same tomographic principles at any region within the bounds of the photoresist. To introduce the R2R CAL concept, we envision tomographic printing in an annulus-shaped region away from the rotation axis. Half of this annulus can be split and "unwrapped". This geometry is immediately recognizable as a roller conveying a strip of material. A differential element of photoresist that travels over the curved portion is exposed to sufficient azimuthal projections, because, theoretically, CAL only requires $\Delta\theta = \theta_{max} - \theta_{min} = \pi$ angular coverage to reconstruct any arbitrary geometry, and, indeed, can generate acceptable reconstructions of most geometries with considerably smaller angular coverage (supplementary section S10) [9]. The proposed configuration is compatible with R2R processing; thus, R2R CAL is a technology capable of producing 3D structures in a continuous strip of photoresist (Figure 1A).

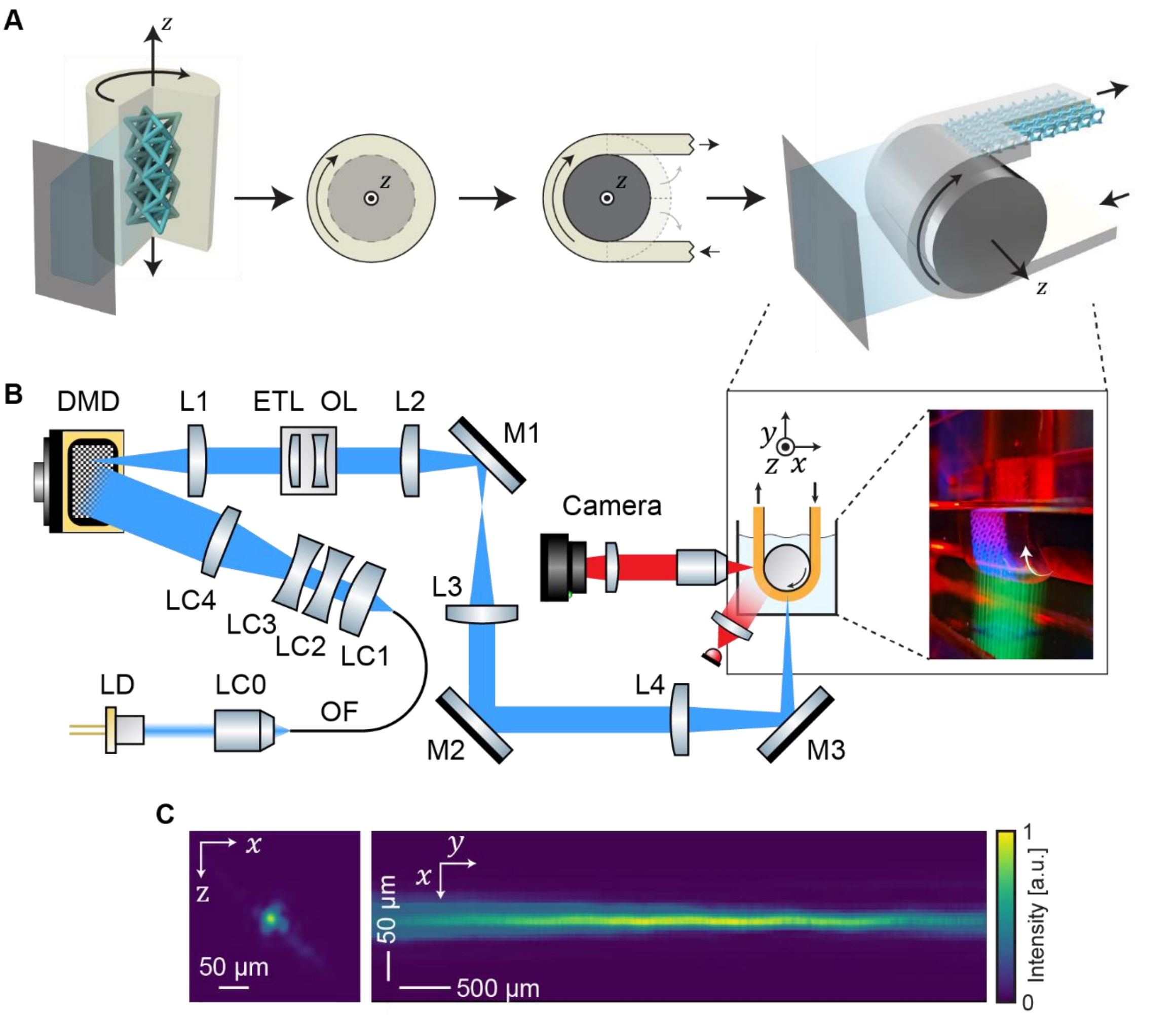

Figure 1. A) The concept of R2R CAL originates from the observation that tomographic printing is possible by unwrapping an annulus from conventional CAL to a R2R configuration. B) The optical system begins with a fiber-coupled laser diode (LD, 442 nm). The output of the optical fiber (OF) is collimated and reflected from the digital micromirror device (DMD). The patterned illumination proceeds through two 4f relays. In the first relay, the electrically tunable lens (ETL), which enables depth-multiplexed projection, is positioned at the Fourier plane. The second relay expands and projects the modulated light pattern onto the photoresist supported by the exposure roller. The inset shows a snapshot taken during printing where the projected light is visible in the immersion medium and the printed structure is visible. Fluorescent dye was added to the immersion medium to enhance the visibility of the projected pattern. C) $x$–$z$ cross section and $x$–$y$ cross section of the projection single-pixel point spread function. Note the coordinate system is shown in B.

*Numerical light propagation model*

In conventional CAL, the target and projection domains (or tomogram and sinogram space, respectively, as referred to in computed tomography literature) are the paired spaces where the virtual model to be printed and the digital images projected to photopolymerize the model reside, respectively. Exposure dosage from simulated backprojections is accumulated in the tomogram domain and compared directly to the prescribed target model and dose. In R2R CAL, however, there is a minor yet important difference. The tomogram domain is instead represented in two different coordinate systems: wrapped ($W$) which corresponds to the geometric configuration where the photoresist is in contact with the face of the cylindrical support; and unwrapped ($UW$) which corresponds to the geometric configuration where the photoresist is flat (Figure 2A).

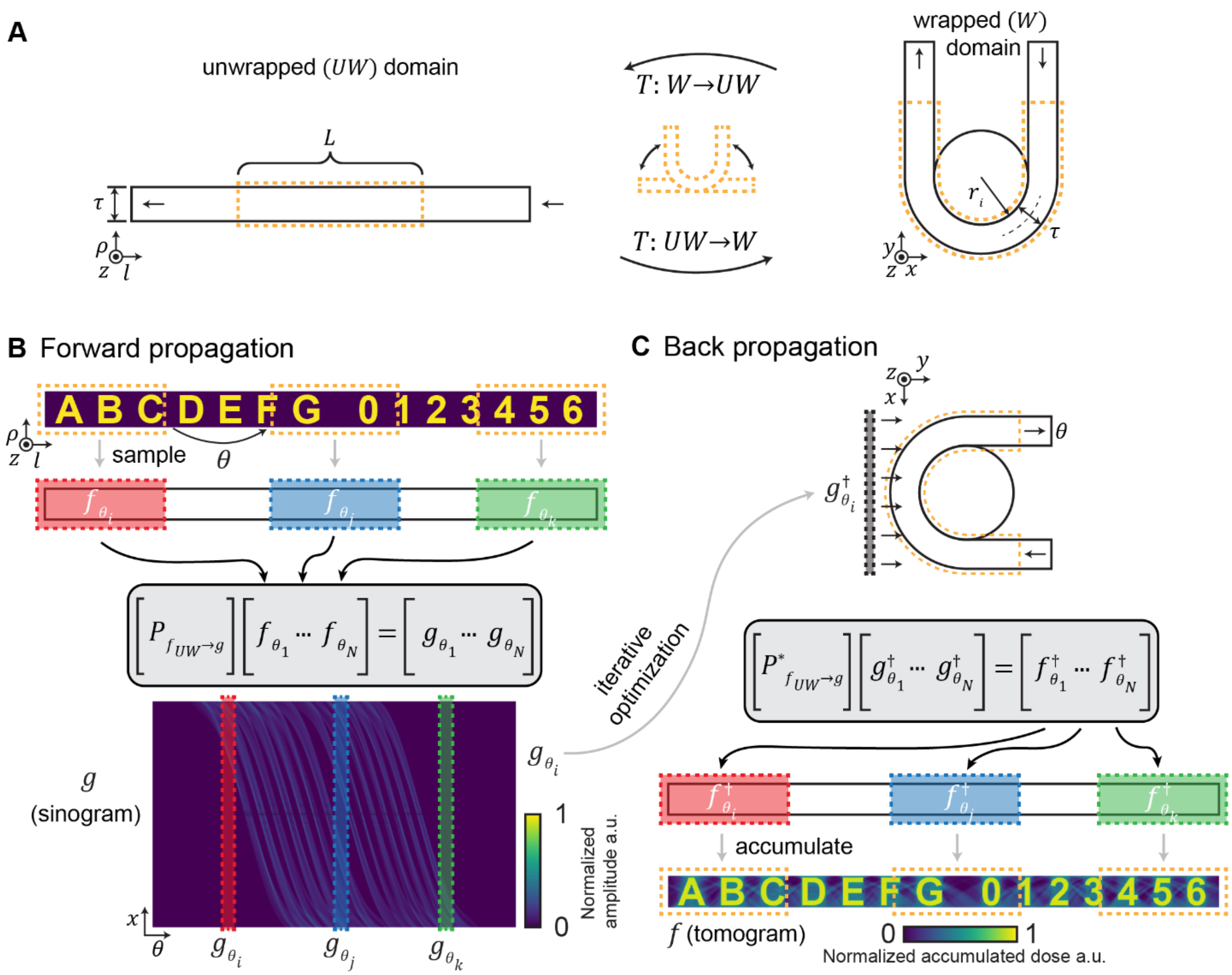

Figure 2. Schematic of propagation in R2R CAL. A) Unwrapped and wrapped tomogram domains and the transformations between them. B) Forward propagation. The unwrapped domain is sampled by a shifting-window operation into blocks $\underline{\boldsymbol{f}}_{\theta_1}$ ... $\underline{\boldsymbol{f}}_{\theta_N}$. The composite propagation is performed as a matrix–matrix multiplication of these blocks by $\underline{\underline{\boldsymbol{P}}}_{f_{UW}\to g}$ which is composed of a transformation $T: UW \to W$ and the forward propagation in the $W$ domain. The result is direct propagation from the tomogram in the $UW$ domain to sinogram $\underline{\boldsymbol{g}}_{\theta_1}$ ... $\underline{\boldsymbol{g}}_{\theta_N}$. C) Back propagation. Physical light propagation (projection of sinogram data $\underline{\boldsymbol{g}}_{\theta_1}$ ... $\underline{\boldsymbol{g}}_{\theta_N}$ into the photoresist) occurs in the $W$ domain. In order to perform optimization in the $UW$ domain, $\underline{\underline{\boldsymbol{P}}}^*_{f_{UW}\to g}$, which is composed of back propagation and then a transformation $T: W \to UW$, can be used to back propagate directly from the sinogram to blocks $\underline{\boldsymbol{f}}^\dagger_{\theta_1}$ ... $\underline{\boldsymbol{f}}^\dagger_{\theta_N}$. $\underline{\boldsymbol{f}}^\dagger_{\theta_i}$ and $\underline{\boldsymbol{g}}^\dagger_{\theta_i}$ represent corresponding tomogram and sinograms resulting from iterative optimization for purposes of illustration. These blocks are then accumulated into the tomogram in the $UW$ domain. This framework avoids expensive coordinate transformations for each propagation and, thus, is useful for iterative optimization techniques like BCLP used in this work.

Simulated propagation of projected light including refraction, attenuation, and absorption occurs in the $W$-domain. Eikonal ray tracing was used to capture the complexities of light propagation. Since the numerical light propagation model is used in iterative optimization, ray tracing for every propagation is computationally expensive. In this work, we utilize algebraic propagation to remove the need to simulate for every propagation. Dose influence coefficients which represent the effect of light exposure on each voxel in the tomogram domain are calculated with ray tracing and stored in the propagation matrix $\underline{\underline{\boldsymbol{P}}}$. In this work, forward propagation is performed algebraically as $\underline{\underline{\boldsymbol{P}}}\underline{\boldsymbol{f}} = \underline{\boldsymbol{g}}$ where $\underline{\boldsymbol{f}}$ is the tomogram, and $\underline{\boldsymbol{g}}$ is the sinogram. Note that, in conventional CAL, $\underline{\underline{\boldsymbol{P}}}$ represents projection from all angles, simultaneously. However, in the R2R CAL system, because the printed object can in principle be limitless in length, the number of projection variables could potentially become unmanageable. A method is therefore needed for reducing the number of projection coefficients.

We leverage the shift invariance in the system to substantially reduce the size of the propagation matrix. The optical path does not change as the web translates; therefore, the propagation matrix can be simplified to only the dose influence coefficients from one angular projection. Specifically, $\underline{\underline{\boldsymbol{P}}}_{f_W \to g}$ is the shift-invariant forward propagation matrix that maps dose influence from the $W$ tomogram domain to sinogram space. In conjunction with a translation (shifting) operator in the $UW$ domain which simulates angular motion in the $W$ domain, we can then repeatedly use these coefficients to compute the overall operation of $\underline{\underline{\boldsymbol{P}}}$. Further, instead of infinite length, we limit the length of the shifting window $L$ based on the penetration depth of light into the straight portion of the web (Figure 2A, supplementary section S11). For this work, $L$ = 1.41 cm. However, $L$ should be adjusted if the absorption coefficient of the photoresist is changed.

However, the prescription dose or intermediate dose during iterative optimization is defined in the $UW$-domain so a coordinate transformation is required to map from the $W$ to the $UW$ domain before $\underline{\underline{\boldsymbol{P}}}_{f_W \to g}$ can be applied. To simplify and reduce the number of operations per propagation, this transformation is subsumed into a composite propagation matrix. Specifically, dose coefficients in the $W$ domain are mapped to the $UW$ domain in the transformation $T: W \to UW$ (shown pictorially in Figure 2A) as follows

$$\underline{\underline{\boldsymbol{P}}}_{f_W \to g}\underline{\boldsymbol{f}}_W = \underline{\boldsymbol{g}}$$

$$T_{W \to UW}\left(\underline{\underline{\boldsymbol{P}}}_{f_W \to g}\right)\underline{\boldsymbol{f}}_{UW} = \underline{\boldsymbol{g}}$$

$$\underline{\underline{\boldsymbol{P}}}_{f_{UW} \to g}\underline{\boldsymbol{f}}_{UW} = \underline{\boldsymbol{g}}.$$

The result is a propagation operator, $\underline{\underline{\boldsymbol{P}}}_{f_{UW} \to g}$, which enables forward propagation from the tomogram in the $UW$ domain to the sinogram with a single matrix–vector multiplication. The adjoint of $\underline{\underline{\boldsymbol{P}}}_{f_{UW} \to g}$, $\underline{\underline{\boldsymbol{P}}}^*_{f_{UW} \to g}$, gives the backpropagation from the sinogram to the optical dose in the UW tomogram domain. Furthermore, the forward propagation matrix, $\underline{\underline{\boldsymbol{P}}}_{f_{UW} \to g}$, and backpropagation matrix, $\underline{\underline{\boldsymbol{P}}}^*_{f_{UW} \to g}$, can be reused along the $z$-axis since each slice is independent under the assumption of small etendue and that propagating rays remain within a plane.

Since $\underline{\underline{\boldsymbol{P}}}_{f_{UW} \to g}$ represents propagation from a fixed angle, web motion (or correspondingly, change in target relative to the projection vector) is expressed as change in the tomogram domain, where $\underline{\boldsymbol{f}}_{\theta_i}$ is a view of the tomogram at an instant in time or angle. Because the unwrapped domain only covers a subset of a possibly longer tomogram space, blocks of the tomogram are sampled with the shifting-window *unfold* function in PyTorch as $\underline{\boldsymbol{f}}_{\theta_0} \dots \underline{\boldsymbol{f}}_{\theta_N}$. The matrix–matrix product of $\underline{\underline{\boldsymbol{P}}}_{f_{UW} \to g}$ and the matrix formed by $\underline{\boldsymbol{f}}_{\theta_0} \dots \underline{\boldsymbol{f}}_{\theta_N}$ performs forward propagation and gives the sinogram for all web positions. The reverse process, back propagation, gives the same sequence of blocks that are accumulated in the tomogram domain with the *fold* function in PyTorch. These forward and back propagation models form the foundation for iterative optimization by gradient descent or other methods [17–20]. The differences between this framework and conventional CAL are further described in supplementary section S2.

*System design*

The optical system was designed based on the microCAL setup in our previous work [12]. A blue laser diode (measured $\lambda_{LD} = 442$ nm center wavelength) was used as the illumination source and coupled into a 100 µm-diameter optical fiber for homogenization. The fiber-coupled source also enabled easy reconfiguration for alignment purposes (supplementary section S6). The pattern displayed on the digital micromirror device (DMD) was projected through the photoresist by two 4f relays. In the first relay with magnification of 1, an electrically tunable lens (ETL) was placed at the Fourier plane. With the ETL, the focus of the projection system can be varied in coordination with image updates, enabling an artificially curved focus surface (see next section). The second relay had magnification of 1.33 which resulted in a maximum projected image size of 27.6 mm × 15.5 mm. The projected single-pixel point spread function (PSF) had a minimum full-width half-maximum of 32.4 µm and depth of focus of 3.8 mm.

For the R2R conveyance system, a polyethylene terephthalate tape 70 mm in width and 75 µm in thickness was used as the web (Figure 3A). The web was wrapped around a series of rollers: an unwind roller controlled by a stepper motor, a series of custom-made idler rollers, a custom-made exposure roller, and a rewind roller driven by an AC torque motor (supplementary section S1). The stepper motor was used to drive the web at a constant rate velocity and the torque motor was used to supply a constant tension and rewind the web. On the tensioned web, gel photoresist which we demonstrated in previous work was cast into films with thickness $\tau = 3$ mm [21]. Using real-time FTIR, we measured conversion as a function of exposure time and observed a sharp conversion increase after a brief induction period. This result indicated that the photoresist had high conversion contrast and would perform well in tomographic printing.

When printing, the exposure roller was submerged in a glass box filled with immersion medium to reduce refraction at the surface of the photoresist and maximize angular coverage (supplementary section S10 and methods). For the prints shown in this work, water with refractive index $n_{med} = 1.33$ at $\lambda_{LD}$ was used as the immersion medium. This medium gives angular coverage of 2.2 rad for photoresist refractive index $n_{pr} = 1.49$ at $\lambda_{LD}$. This combination provided high angular coverage while remaining practical and clean since the photoresist and web are both hydrophobic. While higher refractive index immersion medium was used during testing, since it was highly wetting to the web and photoresist, the burden of additional cleaning and post-processing outweighed the minimal performance improvement gained by increased angular coverage (supplementary section S10).

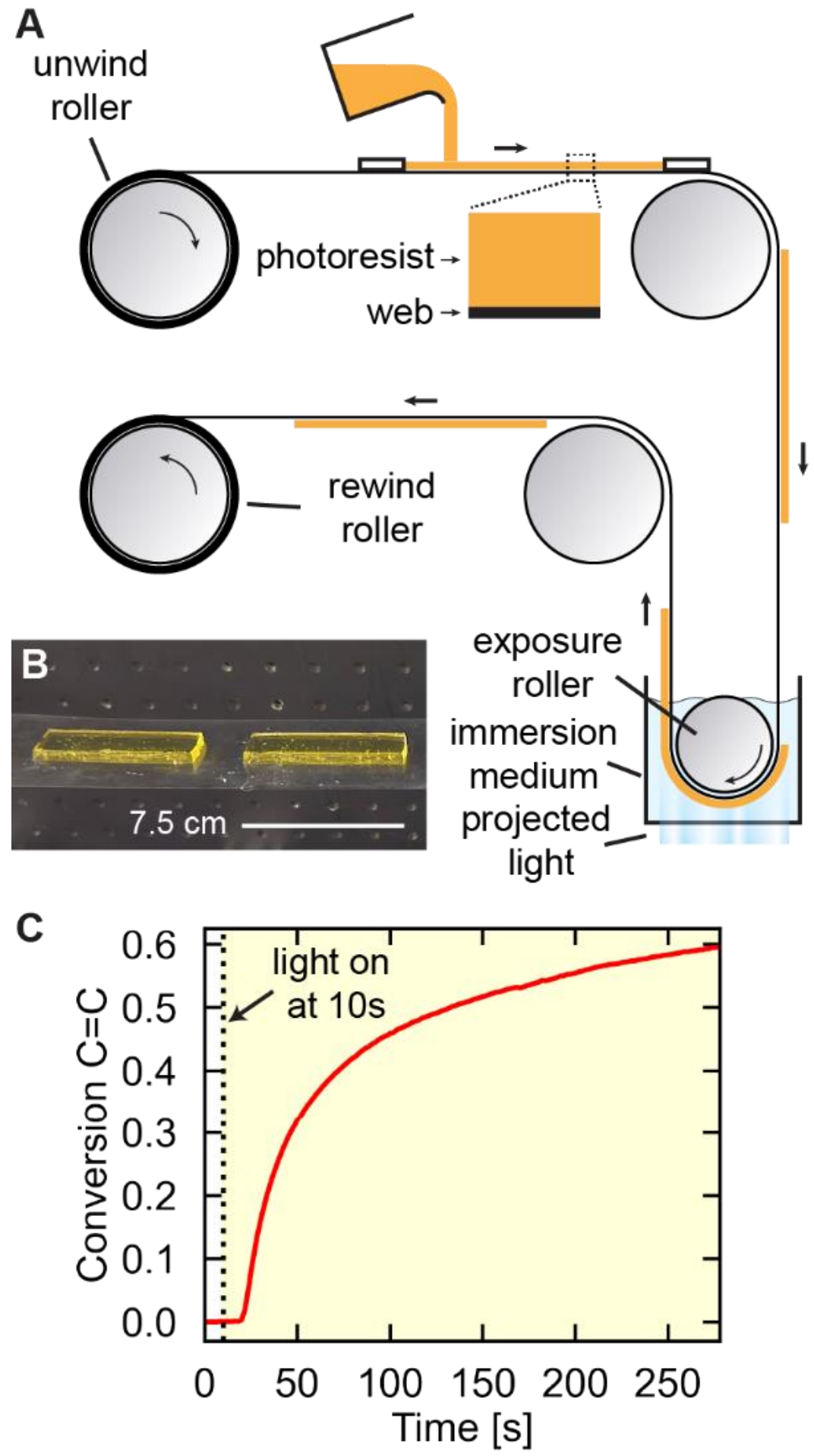

Figure 3. A) R2R conveying system. B) Gel photoresist cast in 3 mm-thick freestanding films. C) Real-time FTIR data of C=C bond conversion.

*Depth- and time-multiplexed projection*

In tomographic 3D printing, high resolution patterning is best achieved when the photoresist is exposed to in-focus light projections for every angle of exposure. Due to the wrapped configuration in R2R CAL, the photoresist traverses a semicircular trajectory. This means that the focus of the optical system must match this shape to achieve the best performance. As it is challenging to design an optical system with a cylindrical focal surface which achieves axially symmetric PSF in addition to telecentricity, we opted instead to utilize time multiplexing to create an artificially prescribed focus shape.

Specifically, an electrically tunable lens (ETL) was added in the Fourier plane of a 1:1 4f relay after the digital micromirror device (DMD). This dynamic element enabled adjustment of the focal length of the projection system along the optical axis in the y-direction (supplementary section S4 and S5). The sum of the radius of the exposure roller and the thickness of the material is well within the total range of focus adjustment. However, to create an artificial cylindrical focus surface, the entire image must not be projected while the focus is adjusted; otherwise regions of the projection image which are not in focus would impinge on the photoresist negating the benefits of focus scanning. Therefore, a depth-binning segmentation scheme was used. The midline of the photoresist layer was used as the targeted focus trajectory. A series of masks was created to block regions of the projection image which would be out of focus for a given ETL power. These masks are shown in Figure 4A with the focus depth overlaid. Since the depth of focus was 3.8 mm, the total depth range from 0 to $-r_i - \frac{\tau}{2} = -1.06$ cm was divided into $N = 3$ depth bins. After masking, the segmented projection images were loaded into the DMD memory and displayed in sync with the ETL drive signal as described in supplementary section S14.

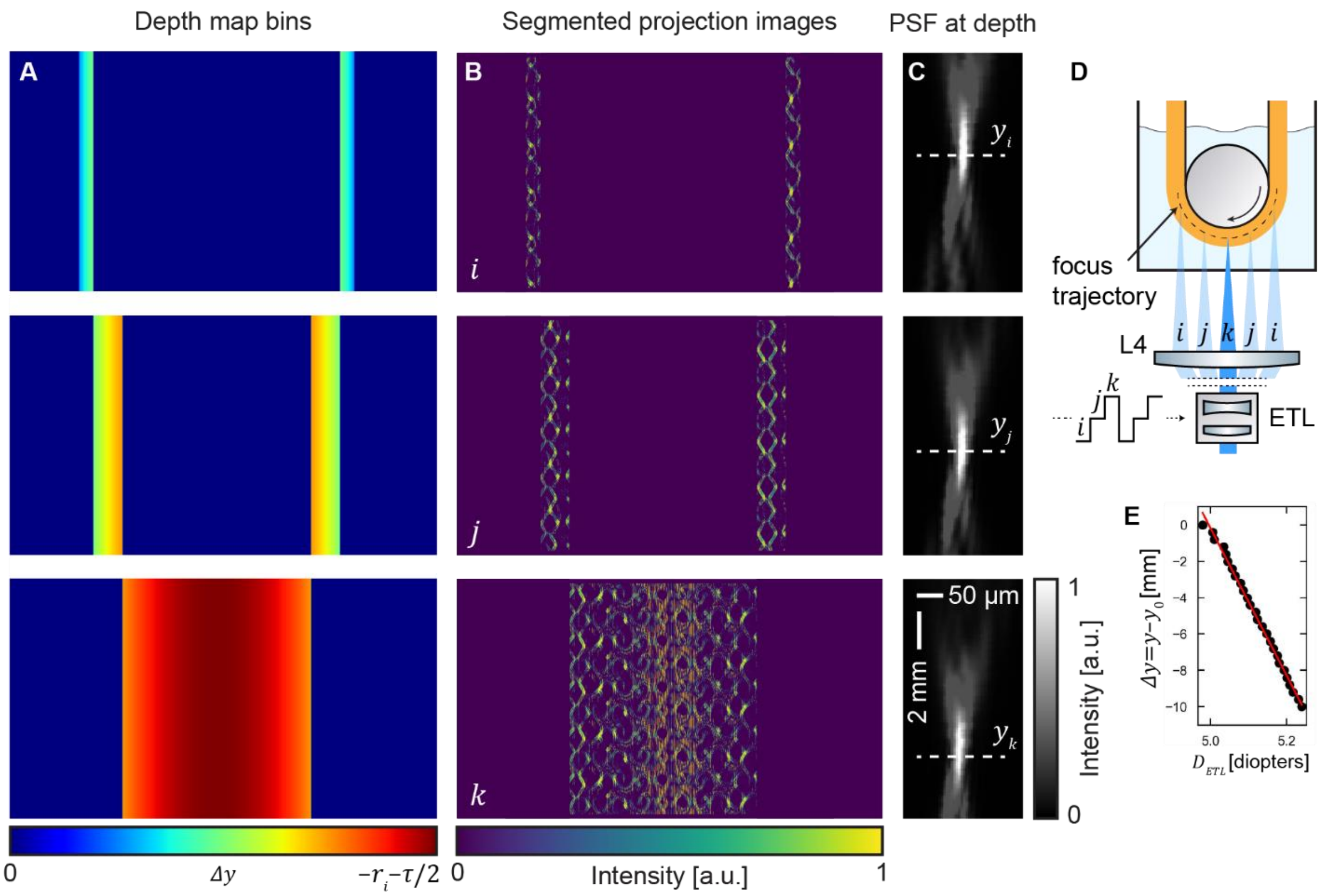

Figure 4. Depth binning for focus scanning A) $N$ depth map bins are applied to a single projection image to generate $N$ depth-segmented projection images. B) Depth-segmented projection images. C) Optical PSF measured at the corresponding depth bins. D) The ETL drive signal is synchronized so that optical power is changed incremented to the power corresponding to a given depth-segmented projection. Note that scales are anisotropic, i.e., scale along the $y$-axis is shortened relative to the $x$-axis. E) Measured focus height vs ETL optical power ($D_{ETL}$).

*R2R CAL printing*

Triply periodic minimal surfaces (TPMS) and 3D lattices are becoming widely used for various applications including filtration, acoustic and optical metamaterials, and biological scaffolding [22–25]. These geometries are challenging to produce with conventional subtractive manufacturing processes due to the complexity and reentrant nature of the positive and negative features. Thus, AM processes have become the standard for manufacturing of these structures. As many AM processes like laser powder bed fusion and stereolithography are limited in size and to batch production, manufacturing of TPMS and lattice-like products at mass scale remains challenging. Here we propose R2R CAL as a scalable technology to fabricate these geometries continuously. An example of such geometry is shown in Figure 5A which has smooth internal surfaces and minimum negative feature sizes smaller than 400 µm. Lattices with minimum feature size of 200 µm and length up to 600 mm were printed. These long, printed constructs can be wrapped upon themselves to potentially form spiral-wound superstructures (see supplementary section S8 and S9).

In addition to connected structures, isolated components can be produced. Since the photoresist is thermally gelled, it acts as support for objects at all stages. For example, this enables overprinting, a mode of VAM where objects can be printed surrounding pre-suspended objects or even liquids. In Figure 5E, a process is illustrated in which doped liquids were injected into the material prior to printing, and capsules were overprinted onto the droplets. This demonstration shows the capability of R2R CAL to print arrays of capsules enclosing previously injected liquids which could be used to produce customized drug carriers with sequenced and/or controllable release rates.

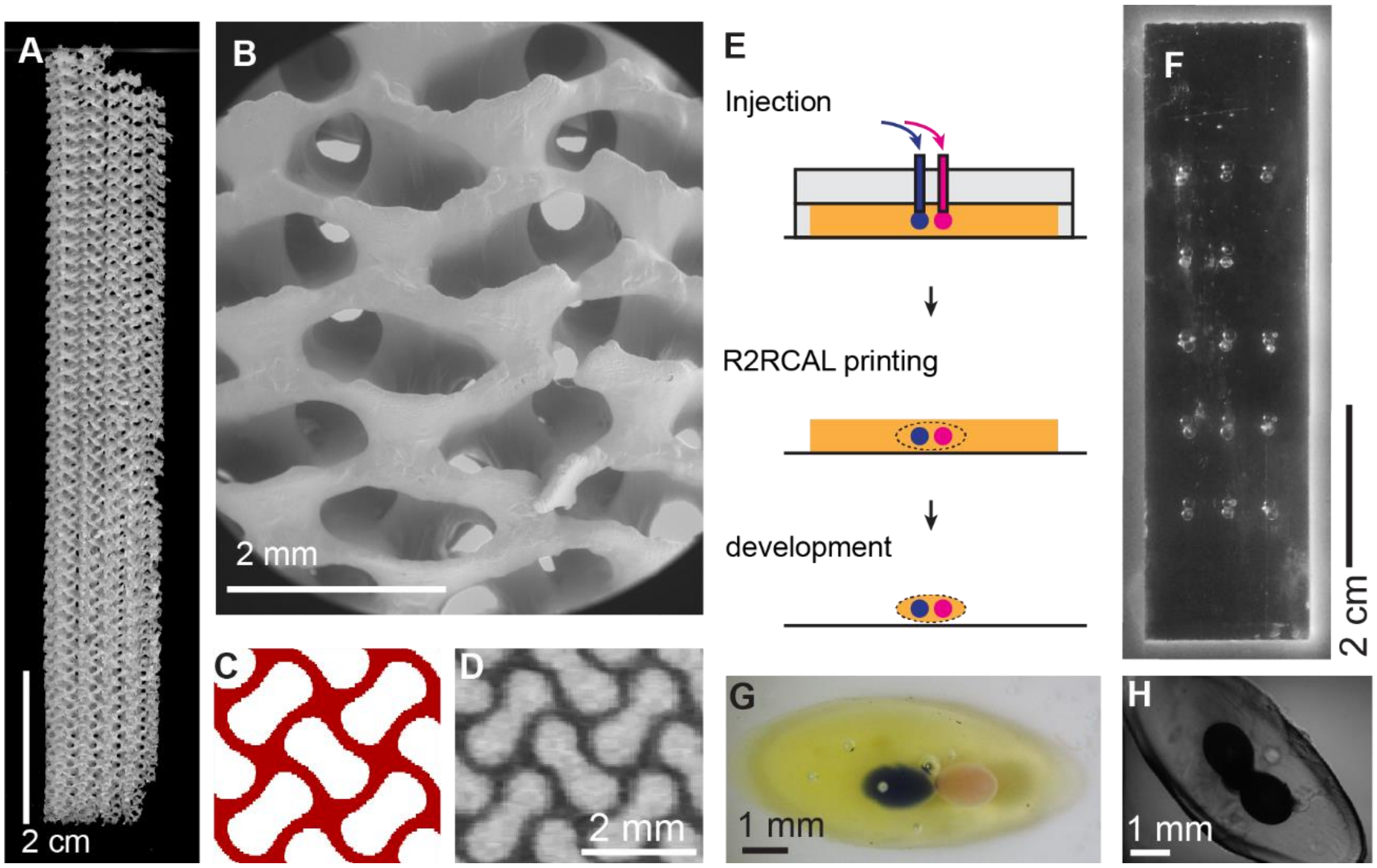

Figure 5. A) R2R CAL-printed sheet gyroid TPMS structure 10 cm in length. B) Scanning electron micrograph shows the internal surface morphology and clear pores. C) Cross-section of the target geometry. White represents void regions and dark (red) represents the target or gel regions. D) Cross-section of an X-ray micro computed tomography scan of the printed structure. Light color represents void regions and dark color represents polymerized material. E) Schematic depicting an overprinting mode of R2R CAL where capsules are printed over doped liquids which are injected into the gel photoresist prior to printing. F) Transmission image of injected water droplets. G) Printed capsules after development. H) Transmission microscope image of the printed capsule.

Four different commonly used TPMS geometries were chosen to demonstrate the fidelity of R2R CAL printing (supplementary section S7). Optimized sinograms were generated by minimizing the band constraint $L_p$-norm (BCLP) objective function using gradient descent [19]. L2 regularization was added to reinforce dose uniformity. After printing and post-processing, the structures were measured with X-ray CT scanning to reveal the complex internal structure (see methods). Micro-CT scans are shown in the bottom row of Figure 6**Error! Reference source not found.**. To evaluate the accuracy of printing, isosurfaces from the micro-CT scans were compared to isosurfaces of the intended voxelized TPMS target geometry (see methods). The mean of Gaussian distributions fit to the signed distance (SD) error distributions were all larger than zero which suggests that all patterns received more than the requisite light dose leading to dilation of features. With refinement of process parameters such as the web speed and laser power, we expect the mean could be brought closer to zero. A significant fraction of the error is likely due to distortion of printed lattice relative to the perfectly periodic and uniform target geometry which would be expected to broaden the distribution. The error within individual cells could be expected to be less than the standard deviation because of this.

Table 1. Statistics of Gaussian distribution fit to signed distance error histograms comparing scanned physical components with the periodic target geometries. The histograms are shown in **Error! Reference source not found.**, second row. μ is the mean, σ is the standard deviation, and $R^2$ is the coefficient of determination of the Gaussian distribution fit.

| Geometry | μ [mm] | σ [mm] | $R^2$ |
|---|---|---|---|
| Sheet gyroid | 0.099 | 0.200 | 0.895 |
| Skeletal gyroid | 0.071 | 0.206 | 0.975 |
| Sheet Schwarz P | 0.142 | 0.229 | 0.893 |
| Skeletal diamond | 0.229 | 0.307 | 0.983 |
| Octet truss | 0.148 | 0.205 | 0.993 |

Overall, coating thickness of 3.00 mm was used, and the maximum projected width is 15.55 mm. The effective printing area, i.e., the area of a $\rho$-$z$ cross-section, was 46.64 $mm^2$. Linear printing rate, i.e., web speed, up to 48.00 mm $min^{-1}$ was achieved which equates to a volumetric printing rate of 2238.93 $mm^3$ $min^{-1}$. The volumetric patterning rate is 45 times greater than our previous work [12]. The printing speed is fundamentally limited by the modulation frequency of the ETL (see supplementary section S12). The printing resolution is fundamentally limited by a combination of the modulation frequency of the ETL and the radial sampling rate of the projection system (the projector pixel pitch along the $x$-axis is in the $W$-domain). Although this is a significant improvement in patterning rate, the width of printed structures is limited by maximum width of the projection. To expand the width and increase prospects of square meter-scale production of 3D structured sheets and membranes, the projection system could be scanned or replicated along the $z$-axis. Alternatively, for structures that are periodic along the $z$-axis, the optical path could be split with beamsplitters or integral image formation using microlens arrays [26]. For a given image projection configuration, speed can be increased by using a source with higher power, increasing the photoinitiator concentration at the expense of higher sensitivity and lower illumination uniformity, or by utilizing a reflective substrate which reflects extraneous light back into the material. With a reflective substrate, simulation of light propagation could become more complex, however, in addition to more effective use of available illumination, it would enable access to projection angles which are normally inaccessible (with a transparent substrate) leading to potentially greater geometric flexibility.

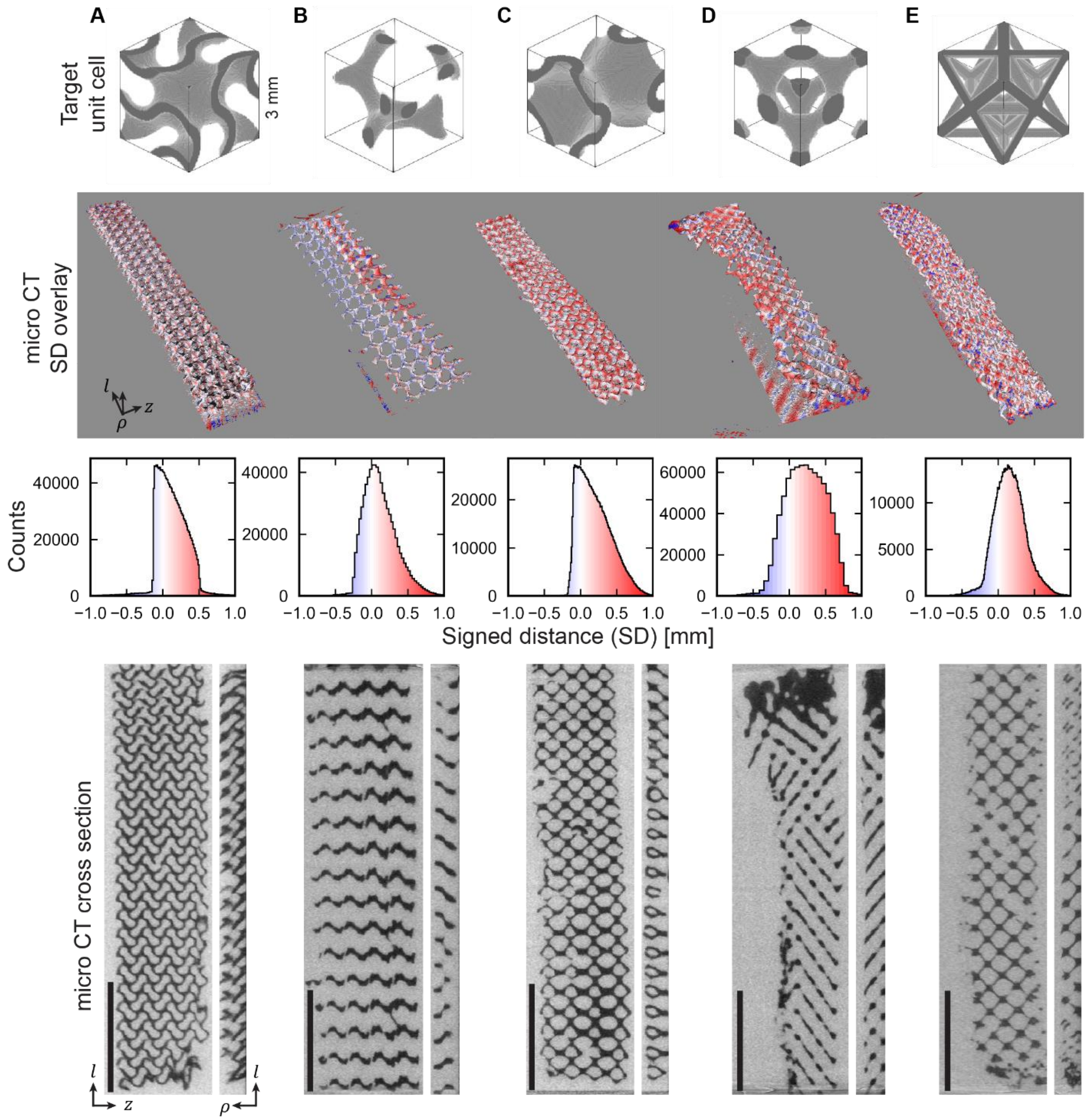

Figure 6. First row: voxel representation of a single unit cell of the target geometry, second row: signed distance colormap overlay on the isosurface of the micro-CT scan of the printed structure, third row: histograms of signed distance error where colors under the histogram correspond to error plotted on the isosurface, bottom row: cross-sectional views of a micro-CT scan of the printed structure. A) sheet gyroid, B) skeletal gyroid, C) sheet Schwarz P, D) skeletal diamond, E) octet truss. Scale bars: 1 cm.

## 3. Conclusion

We have devised and implemented a roll-to-roll continuous tomographic volumetric AM technique which expands the maximum manufacturable part length to a demonstrated 600 mm and towards theoretically infinite size. This technique is supported by a computational framework based on algebraic propagation which enables generation of sinograms for aperiodic geometries longer than the optical simulation volume. The algebraic approach allowed pre-computation of a small propagation matrix which was used to map all sinogram pixels to their contributions to all tomogram voxels. In addition to invariance, the algebraic approach allows us to combine both the optical propagation and transformation between the wrapped and unwrapped material configurations into one composite linear transform. This approach limited the computational burden and enabled optimization of the sinogram for entire structures up to 600 mm long on a computer workstation. We constructed a laser-based optical system with a focus scanning mechanism to fit the wrapped-material R2R exposure configuration and demonstrated printing of complex TPMS structures, lattices, and capsules with minimum positive feature sizes of 200 µm. This work establishes the capability of the continuous R2R CAL method, which could be utilized for various emerging applications including 3D optimized filtration membranes and biological constructs with large cross-sectional dimensions and microscale scaffold features or encapsulated organoids.

## 4. Materials and Methods

*Materials*

Trimethylolpropane triacrylate (TMPTA) was purchased from Fisher Scientific. Camphorquinone (CQ) photoinitiator and ethyl cellulose (EC) were purchased from Sigma Aldrich. All chemicals were used without further purification.

TMPTA and EC thermoreversible monomer was prepared according to our prior work [21]. To the monomer was added 30 mM CQ.

*Real-time FTIR*

A Thermofisher Nicolet iS50 FTIR spectrometer was used to measure conversion vs exposure. A 75 µm thin film of EC gel photoresist was cast between two glass microscope slides. A custom-built collimated light source constructed from a fiber-coupled LED with peak intensity measured at 443 nm (Thorlabs M455F3) was used to illuminate the sample at an intensity of 3 mW $cm^{-2}$. Absorbance spectra were measured in the range of 3000–4000 $cm^{-1}$ at a sampling interval of 1.66 s, with resolution of 8 $cm^{-1}$, and averaging of three spectra. To analyze conversion, the change in absorbance at the peak near the second harmonic of the carbon–carbon double bond (3100 $cm^{-1}$) normalized to the initial absorbance was recorded over the duration of the experiment.

*Printing*

For projection optimization, the BCLP objective [19] was used with $p = 3$, tolerance $\varepsilon = 0.01$ for in-target voxels, $\varepsilon = 0.05$ for out-of-target voxels, logistic response model steepness $B = 10$, momentum $\gamma = 0.5$, uniform L2 regularization coefficient $\lambda_{reg} = 0.04$, and 100 iterations (supplementary section S3). The refractive index in the photoresist was set to $n_{pr} = 1.49$ and outside of the photoresist to $n_{med} = 1.33$. The absorption coefficient was set to $\alpha = 2.2$ $cm^{-1}$ (as measured by spectrophotometry).

EC gel photoresist was cast onto the web using a laser-cut mold made from 3 mm acrylic sheet that was temporarily affixed to the web. The gel photoresist was heated until it melted and then poured into the mold. The mold resulted in cast films 3 mm in thickness and 20 mm in width. The length of the casting was varied depending on the total length of the object to be printed.

Prior to printing, the photoresist was moved to the exposure roller. Here, the projection image sequence, web conveyance, and laser illumination were started in sync. The print proceeded until the final projection image after which the laser was turned off.

After printing on the R2R CAL system, printed specimens were removed from the web and placed in a large petri dish. Excess photoresist along the borders of the printed structure was cut away and recovered for future prints. The petri dish was heated until the unpolymerized photoresist liquified. The liquified photoresist was discarded. Then polyethylene glycol methyl ether acetate (PGMEA) solvent was added to the petri dish. The dish, including developer and the part was slowly agitated for 15 min by sliding it in a circular path on a benchtop. After, PGMEA was removed, and the part was rinsed once with isopropyl alcohol. Finally, the structure was then removed from the dish and dried in air.

*Micro CT scanning*

A General Electric eXplore Locus micro-CT scanner was used to analyze the 3D morphology of printed specimens. The following settings were used: 80 kV tube potential, 450 μA tube current, 2×2 binning, 0.0294 mm detector element spacing, 0.0461 mm spatial resolution, 16-bit depth, 400 ms exposure time, 325.58 mm source-to-detector distance, and 255.10 mm source-to-subject distance. Flat printed samples were taped to thin foam sheets to maximize image contrast between the polymerized material and the substrate. Tomviz software was used to visualize micro CT data [27].

*Surface error comparison*

Using the open-source mesh comparison software, Cloud Compare, the isosurface of the micro-CT data was first registered to the target isosurface by rough manual registration. Then, automatic registration, based on the iterative closest point algorithm, was used to register the meshes closely. After registration was completed, the signed distance (SD) from the micro-CT point cloud to the target mesh was calculated. A Gaussian distribution was fit to the histogram of SD to measure the statistics of the SD error between printed object and the target geometry.

## 5. Acknowledgements

The authors would like to thank Dr. Michael Wendland at the Berkeley Pre-clinical Imaging Core Facility for help with CT imaging. This project was partially supported by Nanomanufacturing Systems Center, a National Science Foundation Nanosystems Engineering Research Center, under Cooperative Agreement No. EEC-1160494. Any opinions, findings and conclusions or recommendations expressed in this material are those of the author(s) and do not necessarily reflect the views of the National Science Foundation.

## 6. CRediT statement:

**Joseph Toombs:** Conceptualization, Methodology, Software, Investigation, Data Curation, Writing – Original Draft and Review & Editing, Visualization. **Chi Chung Li:** Software, Writing – Review & Editing. **Nour Akiki:** Investigation, Writing – Review & Editing. **Yaxuan Sun:** Investigation, Writing – Review & Editing. **Hayden Taylor:** Conceptualization, Writing – Review & Editing, Supervision, Project Administration, Funding Acquisition.

## Supplementary Materials

*S1. System design*

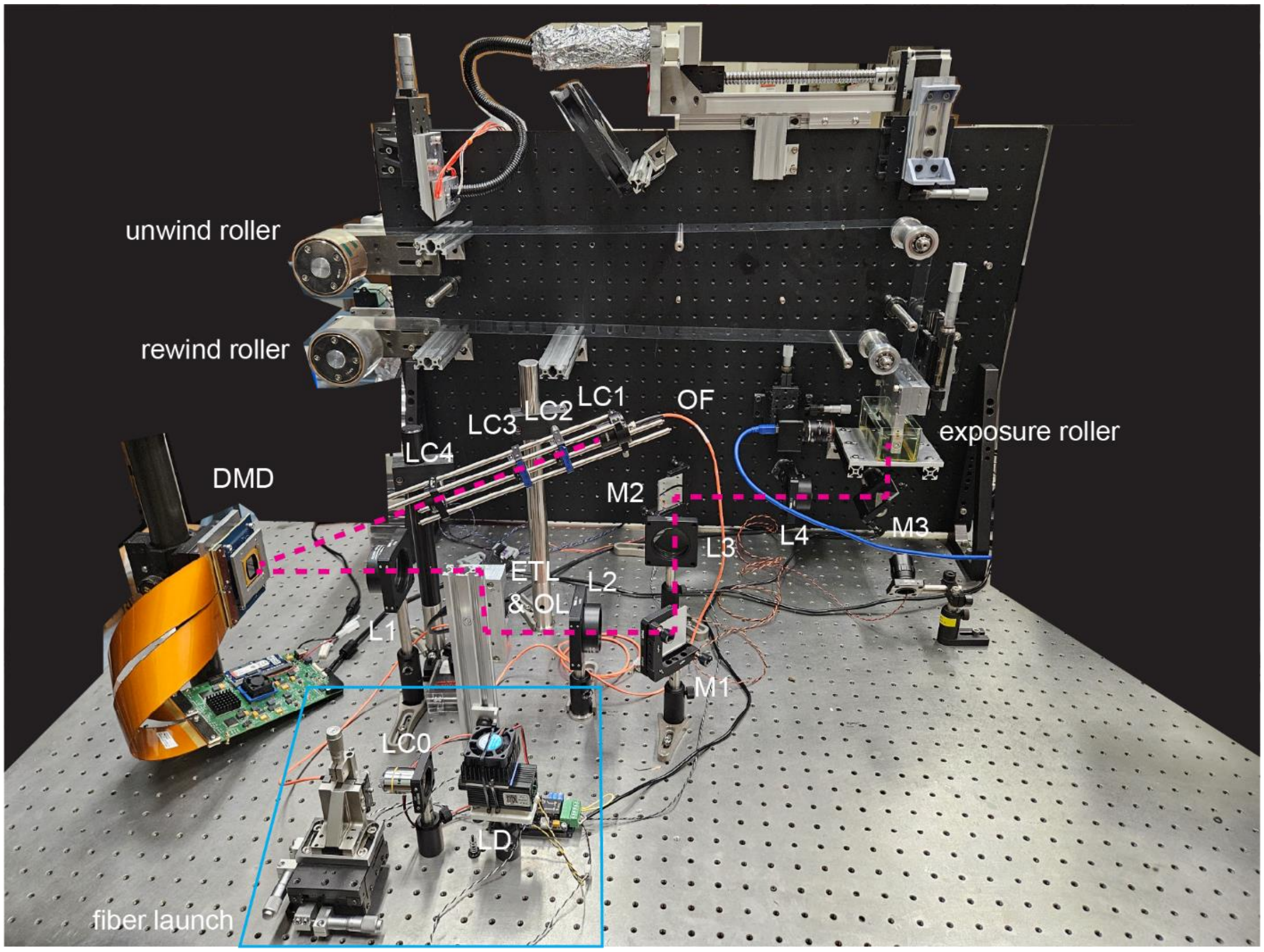

Figure S1. R2R CAL system on optical table. The dashed line represents the optical path after exiting the optical fiber.

Table S1. Optical system components

| Label | Part Number | Focal length (mm) |
|---|---|---|
| LD | Nichia NUBM47 7W | - |
| LC0 | Thorlabs RMS10X | 18 |
| LC1 | Thorlabs AC254-030-A | 30 |
| LC2 | Thorlabs LK1336RM-A | –50 |
| LC3 | Thorlabs LK1743RM-A | –100 |
| LC4 | Thorlabs LA1708-A-ML | 200 |
| DMD | Vialux V9501 with TI DLP9500 DMD | N/A |
| L1 | Thorlabs AC508-150-A-ML | 150 |
| L2 | Thorlabs AC508-150-A-ML | 150 |
| L3 | Thorlabs AC508-150-A-ML | 150 |
| L4 | Thorlabs AC508-200-A-ML | 200 |
| ETL | Optotune EL-10-30-CI-VIS-LD | 100–200 |
| OL | Thorlabs LC4918-A | –200 |
| M1–M3 | Thorlabs ME2S-P01 | N/A |

*S2. Algebraic propagation*

For R2R CAL, algebraic propagation is used in iterative optimization of sinograms. There are important differences between the architecture used in R2R CAL vs conventional CAL. In conventional CAL, the tomogram (the space in which the optical dose response target or intermediate dose response reconstruction (in iterative optimization) is defined) is represented by a single vector $\underline{\boldsymbol{f}}$. The propagation matrix $\underline{\underline{\boldsymbol{P}}}_{f\to g}$ encodes projection at every angle and maps from the $x$–$y$ tomogram domain to the $r$–$\theta$ sinogram domain (for a given $z$-slice). Multiplication of $\underline{\boldsymbol{f}}$ by $\underline{\underline{\boldsymbol{P}}}_{f\to g}$ gives a sinogram in the form of a single vector $\underline{\boldsymbol{g}}$.

For R2R CAL, since the object to be printed can be extremely long, the above computation problem becomes extremely large. To perform optimization efficiently, our approach leverages the shift-invariance property of the system and reuses the invariant part of the propagation matrix as much as possible. In this approach, the tomogram is not represented by a single long vector. Instead, it is represented by $N_\theta$ vectors of finite size that represent the optical dose response target or intermediate dose response reconstruction at different times (or equivalently, angle, since the material moves at constant velocity around the roller) in the unwrapped domain $\underline{\boldsymbol{f}}_{\theta_1} \ldots \underline{\boldsymbol{f}}_{\theta_N}$. Thus, the tomogram is actually a matrix composed of $N_\theta$ columns. The propagation matrix $\underline{\underline{\boldsymbol{P}}}_{f_{UW}\to g}$ encodes projection at only a single angle and maps from $\rho$-$l$ unwrapped tomogram domain to the $r$–$\theta$ sinogram domain . Consequently, multiplication of the tomogram matrix by $\underline{\underline{\boldsymbol{P}}}_{f_{UW}\to g}$ yields a series of vectors $\underline{\boldsymbol{g}}_{\theta_1} \ldots \underline{\boldsymbol{g}}_{\theta_N}$ in the form of a matrix which is directly the sinogram.

Table S2. Algebraic propagation in conventional CAL vs R2R CAL. For each matrix or vector, the row and column dimensions are labeled to illustrate the differences in shape. Note: the dimensions are for a 2D tomogram. For reference, $\rho$ and $l$ represent directions along the thickness and length of the web, respectively, in the R2R CAL configuration. $x$ and $y$ directions lie in a $z$-slice in the tomogram domain of the conventional CAL configuration. $r$ and $\theta$ represent the radial and azimuthal directions in the sinogram domain.

| | **Propagation matrix** | **Tomogram** | **Sinogram** |
|---|---|---|---|
| **Conventional CAL** | $N_\theta \cdot N_r \overset{N_x \cdot N_y}{\left[\quad \underline{\underline{\boldsymbol{P}}}_{f\to g} \quad\right]}$ | $N_x \cdot N_y \overset{1}{\left[\underline{\boldsymbol{f}}\right]}$ | $N_r \cdot N_\theta \overset{1}{\left[\underline{\boldsymbol{g}}\right]}$ |
| **R2R CAL** | $N_r \overset{N_\rho \cdot N_l}{\left[\quad \underline{\underline{\boldsymbol{P}}}_{f_{UW}\to g} \quad\right]}$ | $N_\rho \cdot N_l \overset{N_\theta}{\left[\underline{\boldsymbol{f}}_{\theta_1} \quad \cdots \quad \underline{\boldsymbol{f}}_{\theta_N}\right]}$ | $N_r \overset{N_\theta}{\left[\underline{\boldsymbol{g}}_{\theta_1} \quad \cdots \quad \underline{\boldsymbol{g}}_{\theta_N}\right]}$ |

*S3. Band constraint $L_p$-norm optimization*

In this work, the BCLP optimization scheme was used. The loss function is written as in our prior work [19] with an additional L2 regularization term with coefficient $\lambda_{reg}$:

$$\mathcal{L} = \left(\left(\int_V w(\underline{\mathbf{r}}) \,\left|\left|\mathcal{M}(f(\underline{\mathbf{r}})) - f_T(\underline{\mathbf{r}})\right| - \varepsilon(\underline{\mathbf{r}})\right|^p \, d\underline{\mathbf{r}}\right)^{\frac{q}{p}} + \lambda_{reg}\mathcal{M}(f(\underline{\mathbf{r}}))^2. \qquad \text{(S1)}$$

The regularization term was added to enforce dose response uniformity within the target region. This regularization was especially useful to improve printability of challenging geometries by increasing the concurrency with which all parts of the structure are polymerized, i.e., minimizing under- and over-cure of target regions.

*S4. Ray transfer analysis of focus scanning*

To compute the effect of the ETL on the focus of the optical system for focus scanning, we use ray transfer matrix analysis. The ray transfer matrix for the projection stage of the optical system shown in Figure 1B, specifically lenses L1 to L4, is

$$[T_{s_B}] = \begin{bmatrix} \frac{f_2 f_4}{f_1 f_3} & \frac{f_1 f_2 f_4}{f_3}\left(\frac{1}{f_{ETL}} + \frac{1}{f_{OL}}\right) - \frac{f_1 f_3}{f_2} - \frac{f_2 f_4}{f_3} \\ 0 & \frac{f_1 f_3}{f_2 f_4} \end{bmatrix}, \qquad \text{(S1)}$$

where $f_n$ is the focal length of lens $L_n$, $f_{ETL}$ is the focal length of the ETL, and $f_{OL}$ is the focal length of the offset lens for the ETL. These values are listed in Table S1. A marginal ray emanating from the DMD is represented as

$$[H] = \begin{bmatrix} \frac{f_1 f_2 f_4 \theta}{f_3}\left(\frac{1}{f_{ETL}} + \frac{1}{f_{OL}}\right) - \frac{f_1 f_3 \theta}{f_2} + \frac{f_1 f_3 \theta}{f_2 f_4}\left(y_{air} + \frac{y_u}{n_u}\right) \\ \frac{f_1 f_3 \theta}{f_2 f_4} \end{bmatrix}. \qquad \text{(S2)}$$

The focal point of the system is found by setting the $x$-coordinate of $[H]$ equal to zero i.e. $[H] = \begin{bmatrix} x|_{x=0} \\ \theta \end{bmatrix}$. The distance from L4 to the focal point is

$$y(f_{ETL}) = n_u\left[-f_2^2 M^2\left(\frac{1}{f_{ETL}} + \frac{1}{f_{OL}}\right) + f_4 - y_{air}\right] \qquad \text{(S3)}$$

where $M = f_4/f_3$. The system focuses to the nominal focal plane $y_0$ when the optical power of the ETL is minimized, i.e., when $f_{ETL} = f_{ETL,max} = \min_I |f_{ETL}(I)|$ where $I$ is the drive current

$$y_0(f_{ETL,max}) = n_u\left[-f_2^2 M^2\left(\frac{1}{f_{ETL,max}} + \frac{1}{f_{OL}}\right) + f_4 - y_{air}\right]. \qquad \text{(S4)}$$

The axial scanning distance from the nominal focal plane of the system is $\Delta y = y - y_0$ which is equivalent to

$$\Delta y(f_{TL}) = n_u f_2^2 M^2\left(\frac{1}{f_{ETL,max}} - \frac{1}{f_{ETL}}\right) \qquad \text{(S5)}$$

where $\Delta y > 0$ corresponds to focusing after the nominal focal plane and $\Delta y < 0$ before the nominal focal plane.

Taking the derivative of $y(f_{ETL})$ with respect to the change in focal length of the ETL $\frac{dy}{df_{ETL}}$,

$$\frac{dy}{df_{ETL}} = \frac{-n_u f_2^2 M^2}{f_{ETL}^2}. \tag{S6}$$

The derivative of $y(D_{ETL})$ with respect to the change in optical power of the ETL, $D_{ETL}$, in units of diopters

$$\frac{dy}{dD_{ETL}} = \frac{-n_u f_2^2 M^2}{1000 \text{ mm}} \tag{S7}$$

has units of mm/dpt.

Table S3. Rate of change of focus position for different media refractive index for the R2R CAL system

| $n_u$ | $\frac{dy}{dD_{ETL}}$ (mm/dpt) |
|---|---|
| 1.00 | –40.0 |
| 1.33 | –53.2 |
| 1.45 | –58.0 |

*S5. Focus scanning verification*

The rate of change of focus position with respect to optical power of the ETL $\frac{dy}{dD_{ETL}}$ was verified using a custom focus measurement setup. A CMOS camera sensor (Thorlabs DCC1645C 1280 × 1024) was mounted on a motorized translation stage (Thorlabs PT1-Z9) on the optical axis near the nominal focal plane. A static image of a Ronchi grating with half period length of 10 pixels was displayed on the DMD. The camera sensor $y$-position was incremented by 0.5 mm steps. For a given $y$-position, the optical power of the ETL was varied linearly and an image was captured for each optical power setting. Images from the camera sensor were processed to compute the Tenengrad focus measure [28]. To these focus measurements, a Lorentzian curve was fit. The optical power at the peak of the fit curve was taken as the corresponding optical power for the $y$-position. These $(D_{ETL}, y)$ pairs were used to verify equation S7 with $n_u$ = 1.00 by finding the slope of the line of best fit. The slope was –40.6 mm/dpt (Figure 4E) which is 1.5% error from the theoretical $\frac{dy}{dD_{ETL}}$.

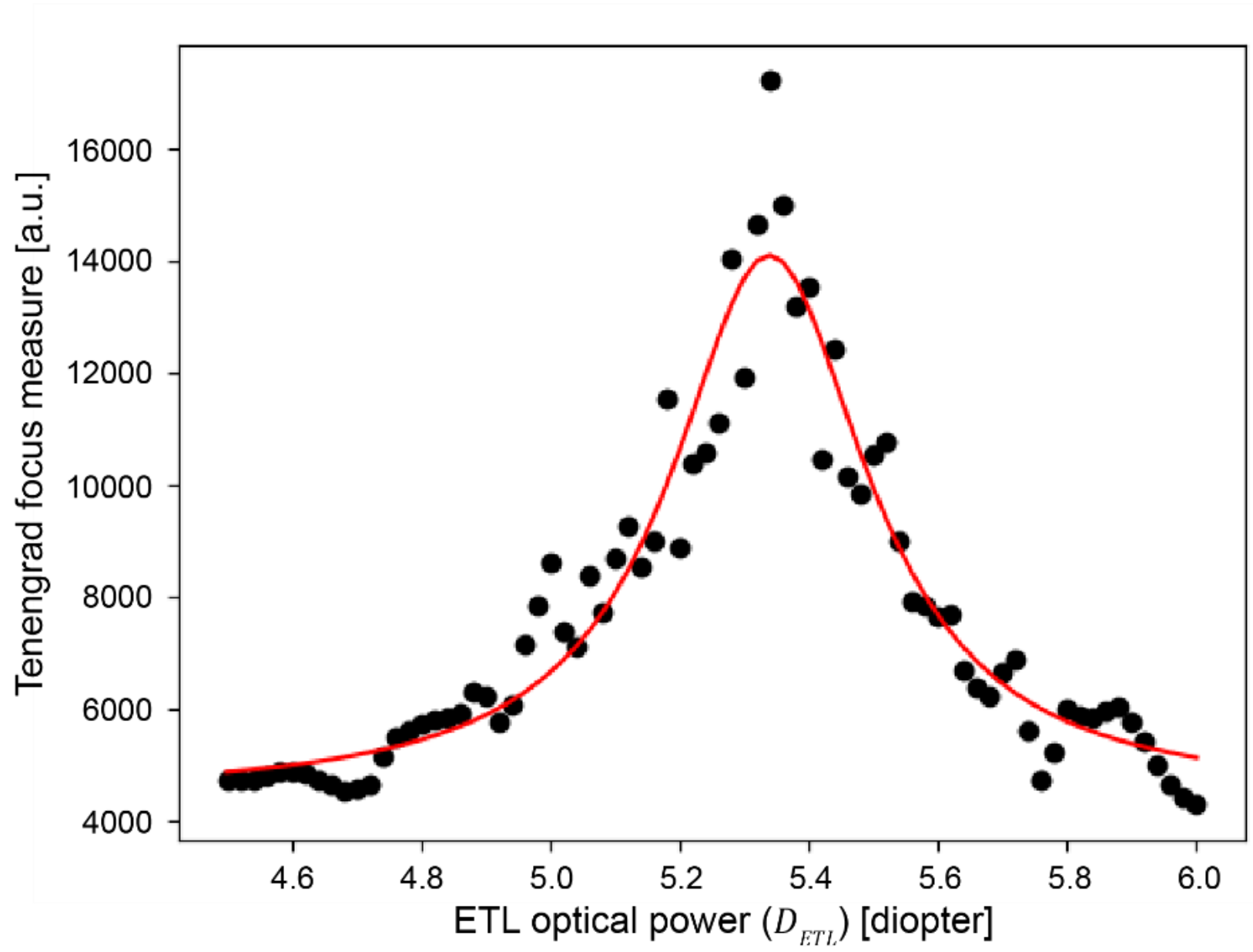

Figure S2. Focus measure vs ETL optical power for a given camera sensor $y$-position. Points represent measurements and the line is a Lorentzian fit to the measurement points.

*S6. Rotation axis alignment*

Accurate rotation axis alignment between the virtual axis of rotation of the projection images and the real roller axis of rotation is crucial to the print quality. Figure S3A–C shows the degradation in reconstruction quality with axis misalignment distance equal to the length of 0, 5, and 10 pixels, respectively. For this work, we use a method which allows repurposing the DMD as a "camera" in conjunction with a photodiode (PD) and reconfiguration of the optical system for alignment with minimal modifications (Figure S3D). The optical fiber end was removed from the collimation assembly and placed above the roller and a collimating lens. The light was collimated and directed towards the roller and backwards into the optical system. In place of the optical fiber in the collimation assembly, a PD is added. In this way, the PD is confocal with the optical fiber and any light that is transmitted backward through the optical system will be focused onto the PD.

A measurement is performed by activating a line of mirrors on the DMD and recording a PD power. After scanning the entire axis of the DMD, a line profile of PD power is extracted. Since the roller blocks the light in this transmission illumination configuration, on the line profile there is a clear transition to black level (Figure S3E). From this profile, the position of the roller in the DMD reference frame could be determined. Finally, with this data, sinogram projection images were offset prior to printing so that the virtual axis of rotation matched this measured axis of rotation.

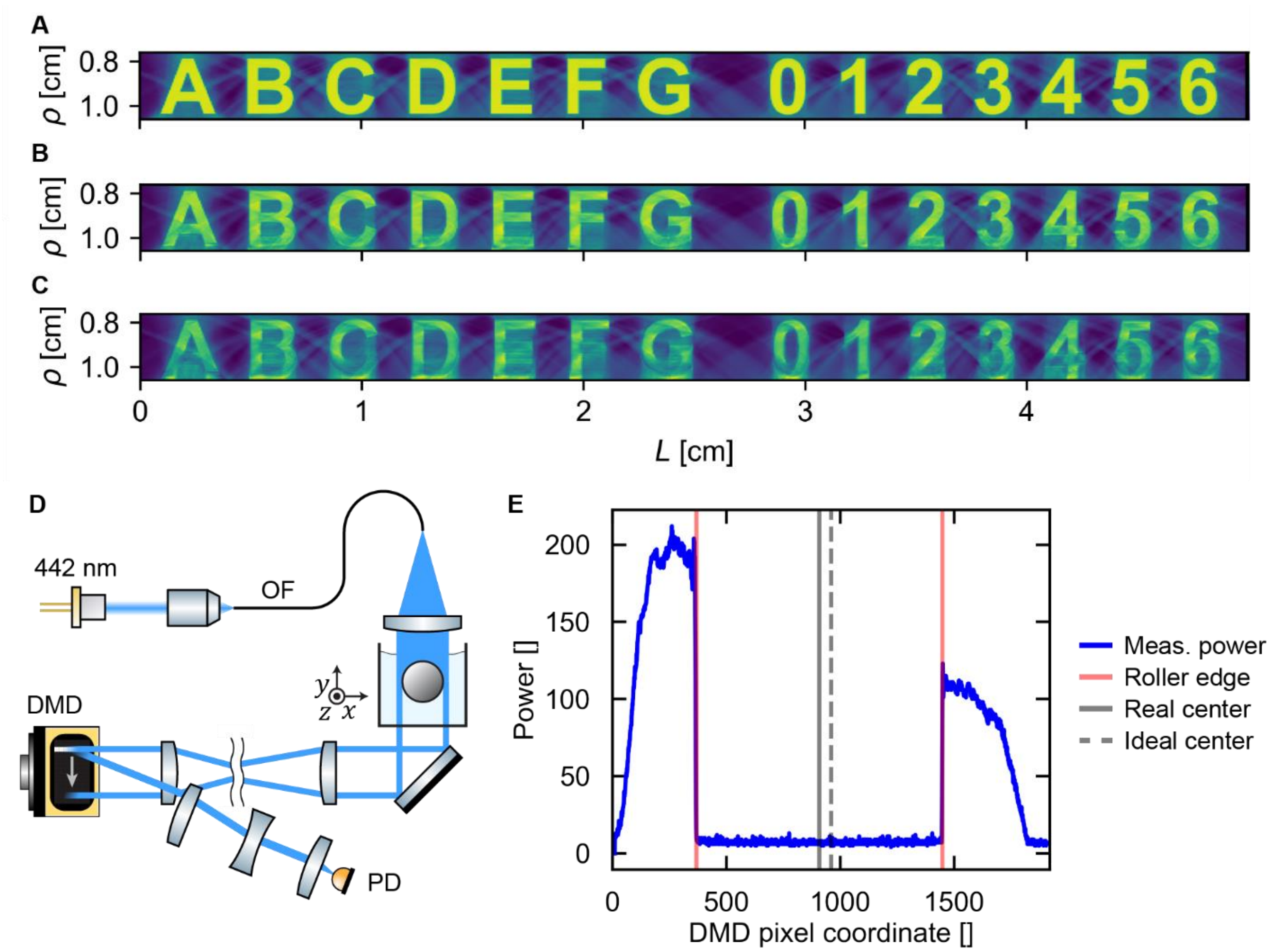

Figure S3. Reconstruction quality for rotation axis misalignment of A) 0 pixels, B) 5 pixels, and C) 10 pixels. Note the pixel size is 0.00144 cm. D) Optical system reconfiguration for alignment. E) PD measurement as a function of the sifting DMD pixel coordinate along the dimension with 1920 pixels.

### *S7. Triply periodic minimal surfaces*

The structures printed in Figure 6 (except the octet truss lattice) are examples of triply periodic minimal surfaces (TPMS). The virtual models of these structures were generated directly from volumetric representations [29]. The trigonometric functions used are given in the table below where $k = 2\pi a$ and $a$ is the unit cell pitch. For all, $a$ = 3 mm.

Table S4. TPMS structure trigonometric functions for volumetric representations

| TPMS type | Function | *c* |
|---|---|---|
| Sheet gyroid | $\phi = (\sin(k\rho)\cos(kl) + \sin(kl)\cos(kz) + \sin(kz)\cos(k\rho))^2 + c^2$ | 0.5 |
| Skeletal gyroid | $\phi = \sin(k\rho)\cos(kl) + \sin(kl)\cos(kz) + \sin(kz)\cos(k\rho) + c$ | 1.2 |
| Sheet Schwarz P | $\phi = \cos(k\rho) + \cos(kl) + \cos(kz) - c$ | 1.0 |
| Skeletal diamond | $\phi = \cos(k\rho)\cos(kl)\cos(kz) - \sin(k\rho)\sin(kl)\sin(kz) + c$ | 0.5 |

*S8. Additional prints*

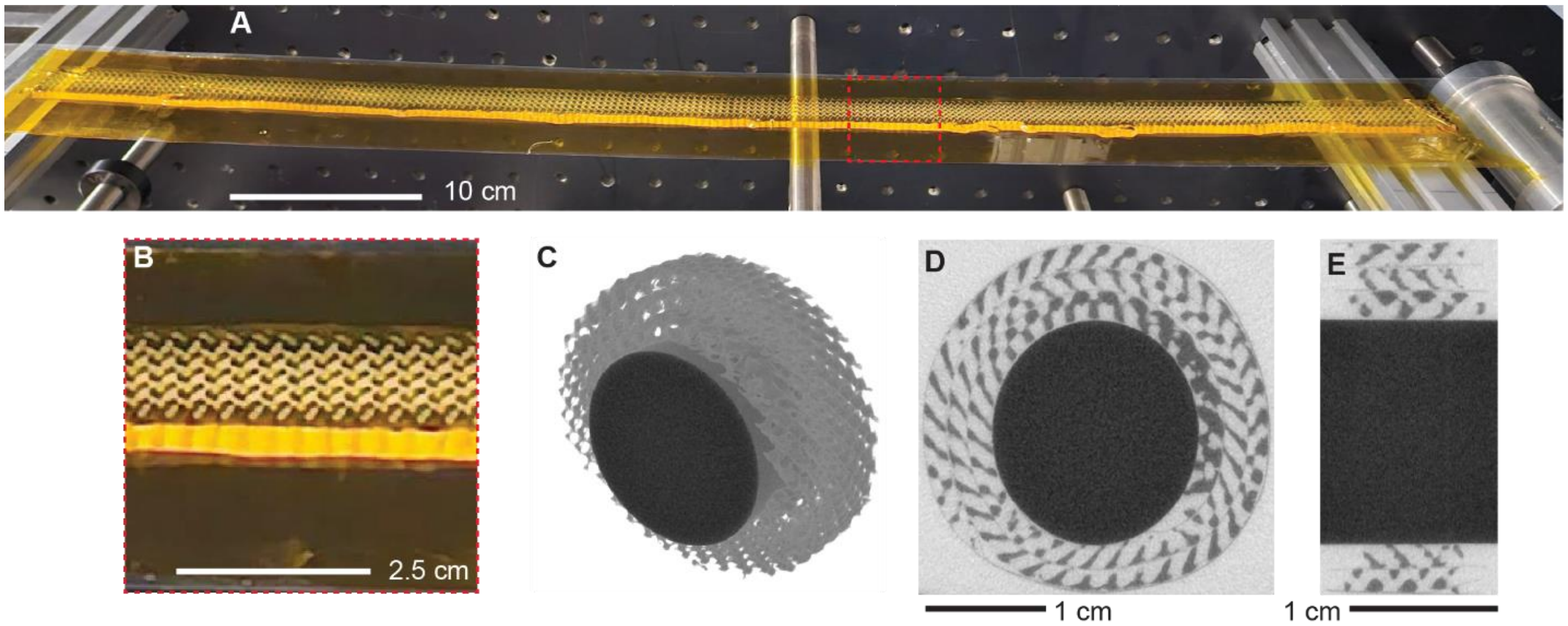

Figure S4. A) Skeletal gyroid structure printed over 600 mm. The structure is not post-processed and remains within the photoresist film. B) Zoom of the dashed region in A. C) Isometric view of a micro-CT scan of the gyroid structure wound on a plastic rod. D, E) Cross-sections of the micro-CT scan.

*S9. Spiral-wound formation of structures*

Although the length of printed structures in the $l$-axis (Figure 2) is theoretically unlimited, the maximum size in the $\rho$-axis is limited by the thickness of the layer of photoresist deposited on the web. To increase the maximum dimension in the $\rho$ direction, the continuous web can be wound on itself or on a central spool. Assuming that adjacent layers of structures can be conjoined after winding, the maximum dimension increases as the number of turns in the spiral-wound structures grows.

In the spiral-wound configuration, we call the resulting form the *composite structure*. If one designs a geometry with dimensions larger than $\tau$, the composite structure must be decomposed, or unwound, into a flat geometry such that it is compatible with the sinogram generation algorithm previously described.

The Archimedean spiral which is commonly used to represent structures wound from materials of constant thickness can be generalized into the algebraic spiral by the equation

$$r = a + k\psi \tag{S8}$$

where $a$ is radius of the starting point of the spiral or the internal wrapping radius, $k$ is the proportionality constant that determines the increase in radius per turn of the spiral, and $\psi$ is the polar winding angle. Specifically, the spiral pitch is $2\pi k$ (Figure S5). The arc length of the spiral is calculated with the formula for the arc length of a polar curve:

$$L = \int_0^{\phi} \sqrt{r^2 + \left(\frac{dr}{d\psi}\right)^2}\, d\psi\,, \tag{S9}$$

Substituting for $r$ gives

$$L = \int_0^{\phi} \sqrt{(a + k\psi)^2 + k^2} d\psi \qquad \text{(S10)}$$

where $\phi = 2\pi N_{turns}$ and $N_{turns}$ is the number of turns in the spiral. For a given set of winding parameters, the total length of the decomposed structure can be calculated by numerical integration of Equation S10.

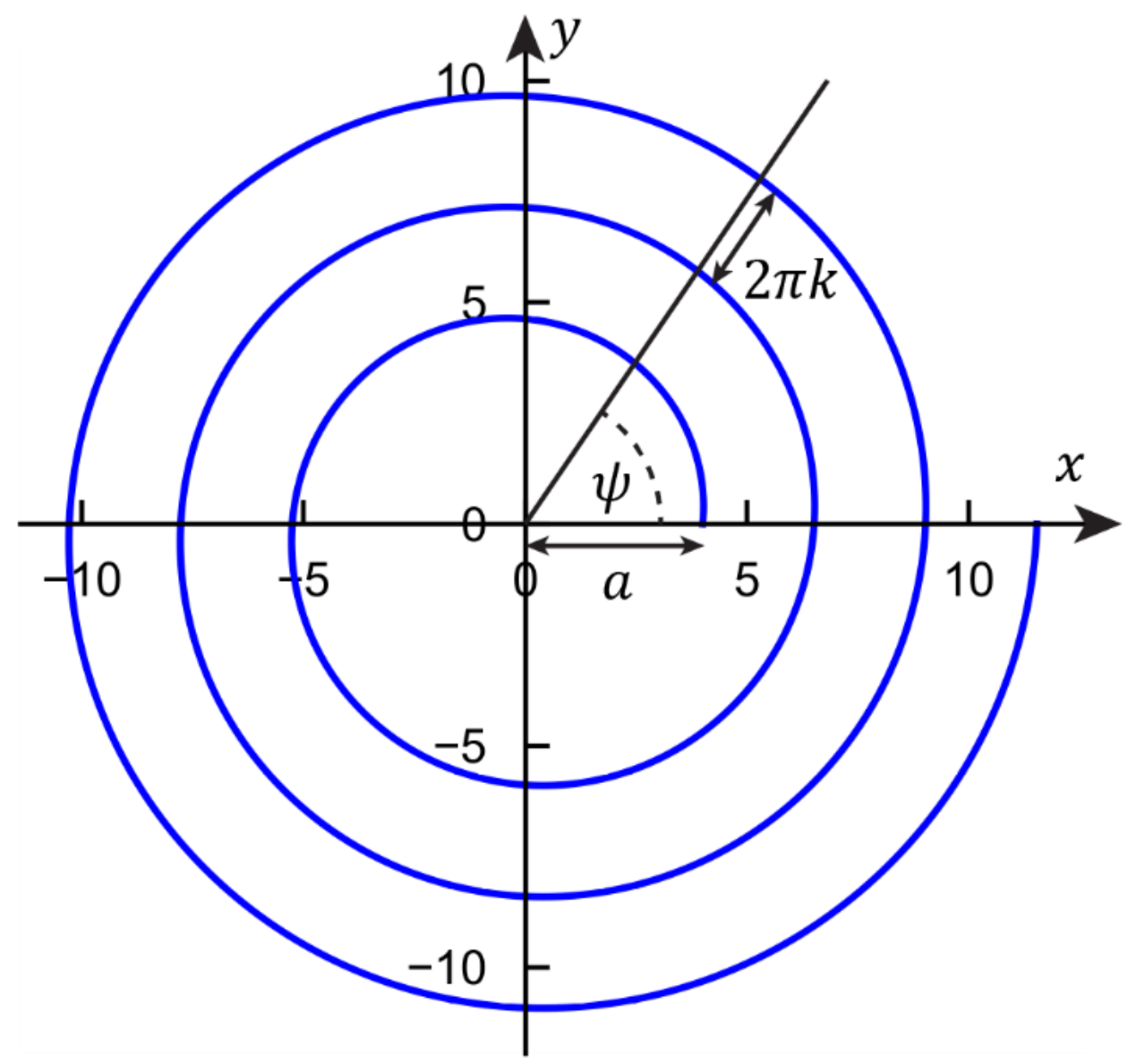

Figure S5. General algebraic spiral

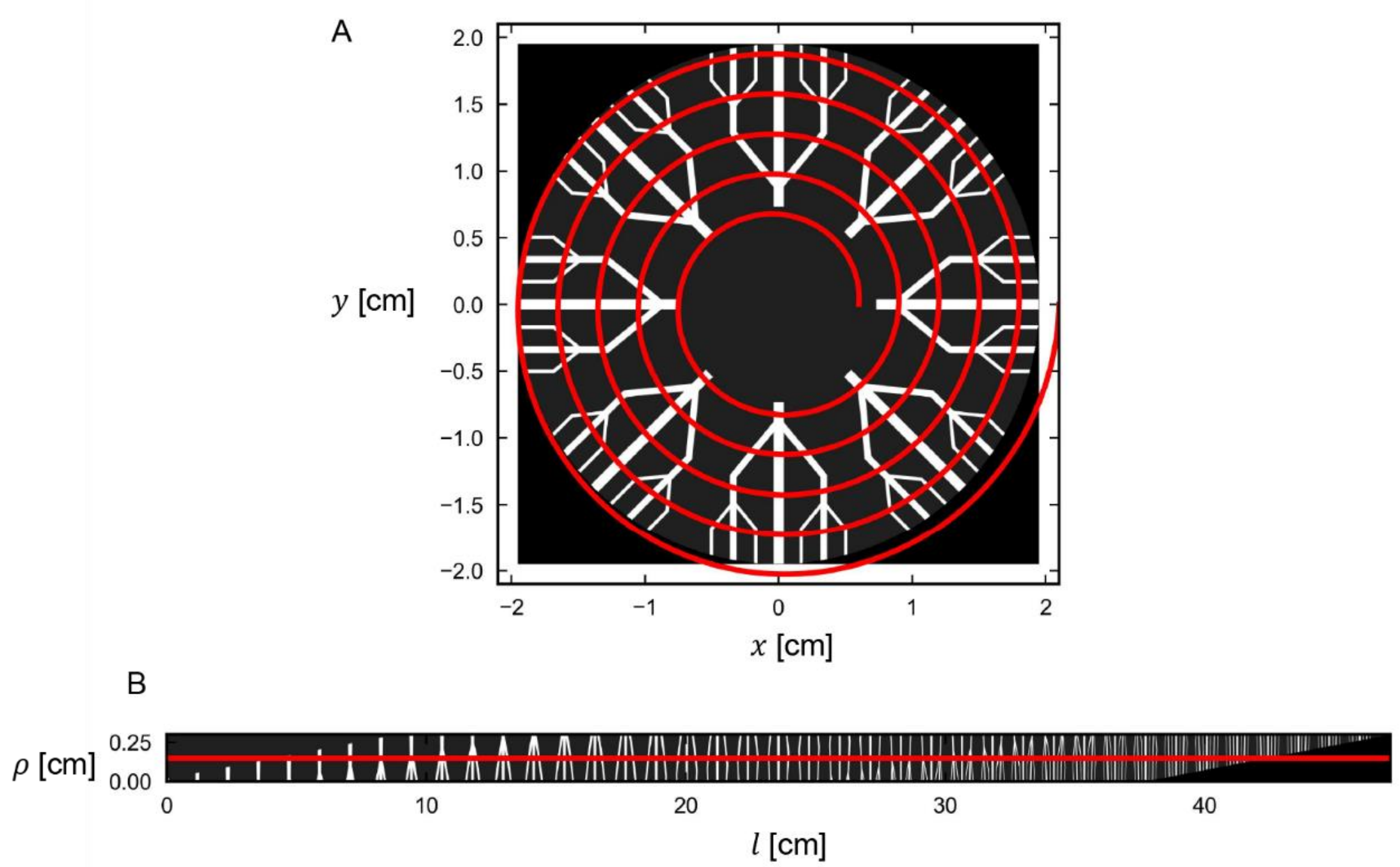

Figure S6. Spiral-wound object decomposition. A) Target object to be decomposed. B) Decomposed and flattened target. The straight red line in B is represented as a spiral in the composite structure in A.

Figure S6 shows an example of composite structure with challenging geometry that spans different feature length scales i.e. small branches at the end of the structure and large total diameter of the bounds. This manufacturing scheme could be useful for biological structures such as vascular geometries with channel diameters and lengths spanning multiple orders of magnitude.

### *S10. Angular coverage*

In computed tomography, angular coverage of the source and detector around the subject to be analyzed is an important concept which governs the fidelity with which the subject can be reconstructed. The central slice theorem states that the Fourier transform of the integral projection through a function is equivalent to a slice through the *k*-space representation of the function orthogonal to the projection vector. Orlov's condition outlines a completeness criterion for sufficient projection data: when the unit vector from source to detector sweeps a line connecting opposite points on the unit sphere, inversion of forward projection is possible [31]. Analogously, complete sampling or coverage of the *k*-space is required to guarantee artifact-free inversion.

In principle, CAL is similar to the inverse of computed tomography. Therefore, the same condition applies to the angular coverage of projection data for CAL. Given sufficiently high angular sampling, physical reconstruction of arbitrary geometries is possible. However, there are a number of process non-idealities that weaken the generality of this assertion. For example, the non-negativity constraint on projection data implies imperfect tomogram contrast and, accordingly, background photoresist exposure which makes certain geometries more difficult to produce.

Theoretically, the semicircular portion of photoresist in the R2R CAL configuration supports complete angular coverage according to Orlov's condition. Although the light source is stationary in space and the photoresist is moving along the semicircular trajectory, the reference frame can be placed on a differential element of photoresist. In this reference frame, the projection vector sweeps a great semicircle source trajectory on Orlov's sphere. However, this is the ideal scenario where there is no refraction at the interface between the surrounding medium and the photoresist. If the medium has a smaller index of refraction than the photoresist, there is refraction at the interface (Figure S7) and the source trajectory arc length (and k-space coverage) is reduced.

Also, there will be nonzero reflectance at the interface which is dependent on the index contrast and angle of incidence. This leads to suppression of the contribution of projections at high angles of incidence. However, in this work, this effect is compensated by applying a multiplicative factor to the projection data which is the inverse of the Fresnel transmittance. Since there is a limited dynamic range and power in the projection system, it is anticipated that this effect and magnitude of compensation required could become important as the printing speed is increased for industrial production. In this work, the effect was not a significant limitation.

In this section, the effect that reduced angular coverage has on the reconstruction fidelity is investigated. The reduction in angular coverage for a given ratio of index of refraction between the medium and the photoresist, $m = \frac{n_{med}}{n_{pr}}$, can be found simply with Snell's law as

$$\theta_t = \sin^{-1}(m \sin \theta_i) \tag{S11}$$

where $\theta_i \in \left[-\frac{\pi}{2}, \frac{\pi}{2}\right]$ is the incidence ray angle and $\theta_t$ is the transmitted ray angle. In *k*-space, the angular coverage reduction results in a missing wedge. The limited angle problem is commonly encountered in computed tomography configurations where the source and detector trajectories are physically constrained by the geometry of the subject [32]. In particular, a closely related configuration to R2R CAL is computed laminography where the subject has a large sheet-like form which makes the conventional computed tomography trajectory challenging or leads to excessive absorption along certain angles [33]. Several approaches have been developed to address the limited angle or missing wedge problem including Fourier space or sinogram extrapolation/interpolation or sinogram completion with neural networks [32,34]. However, in practice, reconstruction is physical: i.e., missing projection data cannot be synthesized and are simply absent if the trajectory does not permit it. Therefore, it is imperative to maximize the angular coverage or, otherwise, to optimize the angular coverage for a particular target geometry.

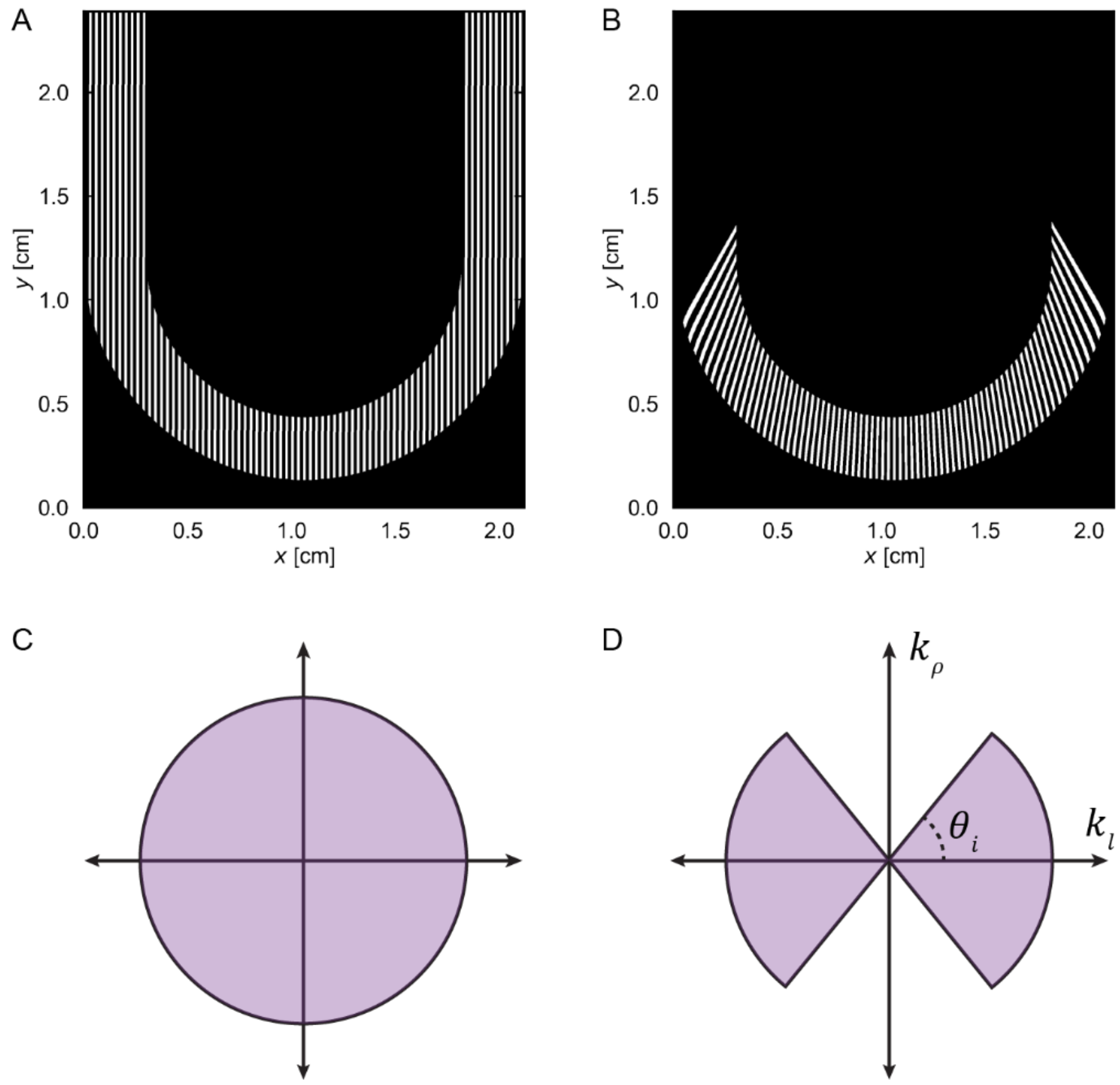

Figure S7. Backprojected ray diagrams for A) $m$ = 1.00 and B) $m$ = 0.75 and *k*-space coverage for C) $m$ = 1.00 and D) $m$ = 0.75. Shaded region represents the portion of *k*-space that is covered.

The effect of angular coverage on the tomographic optical dose PSF was simulated by first assuming a projected pixel beam has Gaussian beam characteristics and accumulating optical dose PSF's from a series of angles within the angular range $\theta_i \in [\theta_{min}, \theta_{max}]$. In Figure S8A, B, the cumulative tomographic optical dose PSF is shown for $m$ = 0.75. The cumulative PSF FWHM along the $\rho$-axis increases as $m$ decreases. However, along the $l$-axis, cumulative PSF FWHM actually decreases as $m$ decreases. This means that there is a (perhaps counterintuitive) potential benefit to using a R2R CAL configuration with smaller angular range.

Furthermore, it is possible to generate a sinogram for geometries which have most spatial frequency content within the angular coverage for a given $m$. In Figure S9, sinograms were optimized for the prescribed dose shown (with randomized structure and broadly distributed spatial frequency content) for several values of $m$ in the range $0.67 \leq m \leq 1.00$. Additionally, a metric was created to compare the dose response contrast:

$$\xi = \frac{\int_Q \mathcal{M}\big(f(\underline{\mathbf{r}})\big)\, d\underline{\mathbf{r}}}{\int_V \mathcal{M}\big(f(\underline{\mathbf{r}})\big)\, d\underline{\mathbf{r}}} \tag{S12}$$

where $V = \{\underline{\boldsymbol{r}} : \underline{\boldsymbol{r}} \in \mathbb{R}^3\}$ is the material simulation domain and $Q = \{\underline{\boldsymbol{r}} : f_T(\underline{\boldsymbol{r}}) > 0\}$ is the set where the target dose response function is greater than zero. $\xi$ is effectively the fraction of the reconstructed dose response function within the target set where $\xi = 1$ is the highest value where all dose is in the target set and $\xi = 0$ is the lowest value where none of the dose is in the target set.

Although the angular coverage decreases as *m* decreases, the optimized reconstruction dose response is largely unchanged between conditions besides an 8.9% decrease in $\xi$. This suggests that reduced angular coverage does not necessarily lead to a substantial compromise in reconstruction quality and that printing in this configuration is viable for a wide range of target geometries. Obviously, there are caveats, for example, a target geometry with high frequency line space patterns along the $\rho$-axis will have spatial frequency content which is not covered in reduced angular coverage conditions. For this type of geometry, reconstruction contrast and printing fidelity would be compromised to a much higher degree. In future systems, a combination of reflectance compensation for large $\theta_i$ and careful design of a diverging projection system, where refraction bends rays into a direction within the photoresist that is tangential to the roller, thus achieving full angular access, i.e., $\theta_t \in \left[-\frac{\pi}{2}, \frac{\pi}{2}\right]$, could enable printing of the most challenging geometries while eliminating the need for an immersion medium.

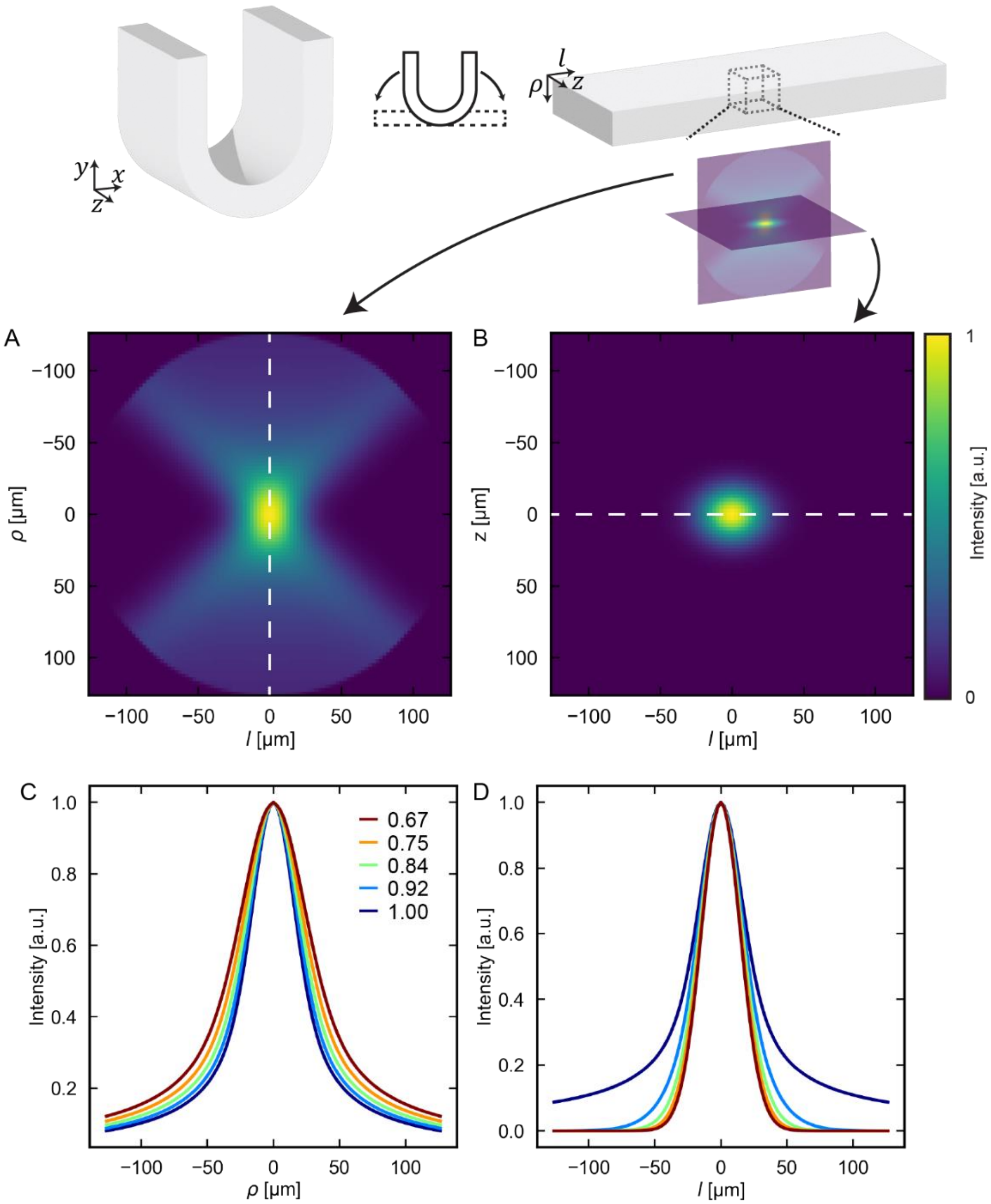

Figure S8. A) $\rho - l$ and B) $l$–z cross section of the cumulative tomographic optical dose PSF for $m$ = 0.75. 1D profiles of cumulative PSF intensity on the C) $\rho$-axis and D) $l$-axis for different refractive index ratios $m$.

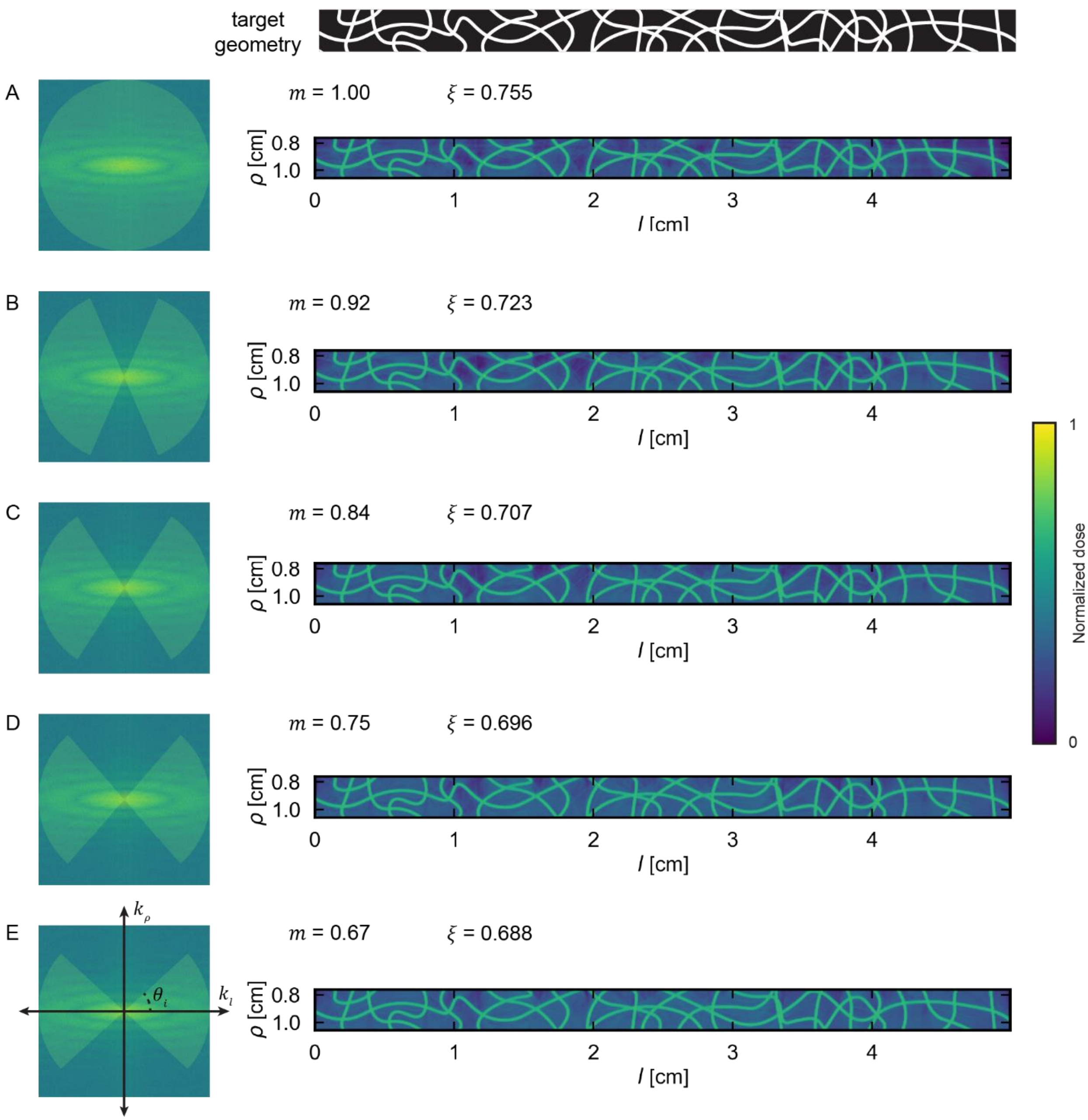

Figure S9. Effect of angular coverage on reconstruction fidelity. The left column shows the spatial frequency coverage for different refractive index ratios. The right column shows optimized dose reconstructions for these different cases with the target prescription dose shown at the top of the column. A) $m$ = 1.00, B) $m$ = 0.92, C) $m$ = 0.84, D) $m$ = 0.75, E) $m$ = 0.67.

## *S11. Absorption*

In the wrapped domain, it is important to simulate sufficient domain size such that ray tracing is not truncated before its energy approaches zero. The photoresist is cast or deposited in the flat state and then it is deformed in the wrapped state during exposure. We assume that the gel is incompressible and thus bending does not cause volume change. From this assumption, the absorption coefficient is considered uniform through the thickness of the material. The optical path length $\beta$ is calculated as

$$\beta = \sqrt{r_o^2 - x^2} + y$$

assuming that rays propagate without refraction at the surface. The origin $(x,y) = (0,0)$ is at the axis of the roller. With the optical path length, the volumetric absorbance factor $\alpha_V$ can be calculated as

$$\alpha_V = \alpha e^{-\alpha\beta} \tag{S15}$$

where $\alpha$ is the absorption coefficient of the material at the actinic wavelength. The absorbed energy per unit volume per unit time is

$$I_{abs} = I_0 \alpha_V \tag{S16}$$

where $I_0$ is the intensity of illumination at the surface of the material. Assuming that intensity $I = I_0 e^{-\alpha\beta} \leq 0.05 I_0$ can be neglected, the minimum simulation length of the straight portion of the web $L_{UW}$ can be found simply by solving for $\beta$ because the maximum penetration depth is expected at the tangent point where $c = r_o$ and $\phi = \frac{\pi}{2}$ and a ray passes through only the unwrapped web. The minimum $L_{UW} = \frac{\ln 0.05}{\alpha}$.

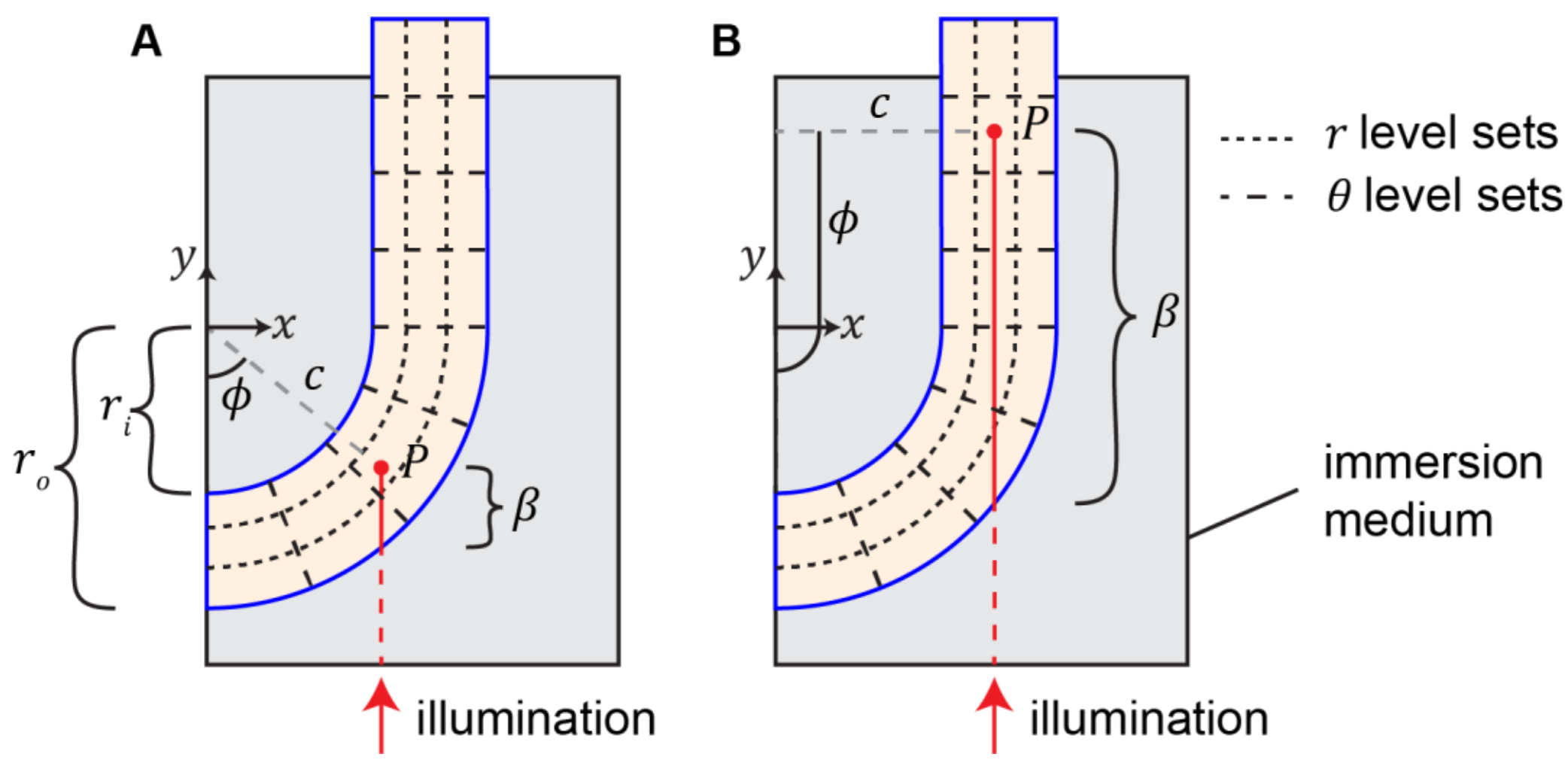

Figure S10. Nomenclature for optical path length.

The absorption coefficient of the EC photoresist (including 30 mM CQ) was measured with a VWR dual beam UV-6300PC spectrophotometer (Figure S11A). The measurement was performed in acrylic 1 cm path-length cuvettes. Isopropyl alcohol was used as a reference. In this case, $\alpha = 2.2$ cm$^{-1}$ and the minimum $L_{UW} = 1.36$ cm. Here we use $\frac{\pi(r_i+r_o)}{4} = 1.41$ cm which is equivalent to one quarter of the circumference of the material wrapped in a circle.

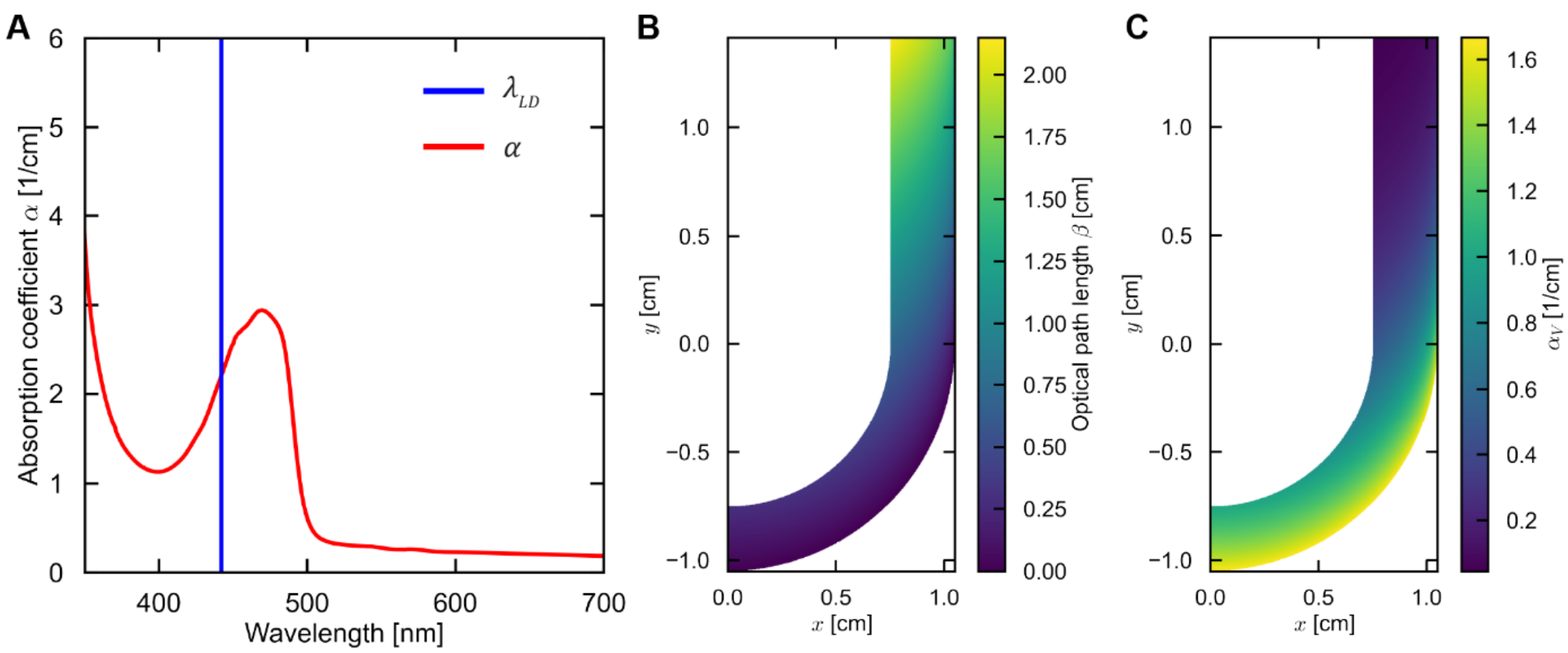

Figure S11. A) Absorption spectrum of EC photoresist. The vertical line indicates the wavelength of the laser diode. B) The optical pathlength $\beta$ in the wrapped domain. C) Volumetric absorbance factor $\alpha_V$ in the wrapped domain.

### *S12. Tangential sampling rate*

The spatial sampling rate in the radial direction is fixed by the optical system. However, the tangential sampling rate is limited by a combination of factors including the photoresist velocity (or web speed) $v_{pr}$, number of focus scanning depth bins $M$, roller radius $r_i$, photoresist thickness $\tau$, angular motion per DMD cycle $\delta\theta$, frequency of the DMD $f_{DMD}$, and the frequency of the ETL $f_{ETL}$. Directly, the tangential spatial sampling rate is limited by

$$f_{tangential} \leq \frac{f_{DMD}}{v_{pr}} \tag{S17}$$

where

$$f_{tangential} = \frac{1}{\delta\theta\left(r_i + \frac{\tau}{2}\right)} \tag{S18}$$

However, the maximum modulation frequency of the ETL and level of depth binning imposes a constraint on the maximum DMD frequency

$$f_{DMD} \leq \frac{f_{ETL}}{M} \tag{S19}$$

assuming that each angular sample is represented by $M$ images displayed sequentially. The ETL used in this work had a minimum response time of 5 ms (using low-pass filtered drive signals to minimize setting time) and thus a maximum frequency of 200 Hz. On the other hand, the DMD used in this work had a maximum frequency of 266 Hz for 8-bit images. Therefore, it is more appropriate to express the tangential sampling rate in terms of the ETL limitation as

$$f_{tangential} \leq \frac{f_{ETL}}{M v_{pr}}. \tag{S20}$$

Obviously, slower web speed results in higher maximum $f_{tangential}$. As an example, for $v_{pr} = 1$ mm s$^{-1}$, the tangential sampling rate is limited to a maximum of 66 mm$^{-1}$ for $M = 3$ depth bins.

### *S13. Photoresist flow*

From the Navier-Stokes equation, the velocity of a fluid film down an inclined surface under gravity-driven flow and a no-slip boundary condition is

$$u(\hat{y}) = \frac{\rho g \sin\alpha}{\eta}\left(\hat{y}\tau - \frac{\hat{y}^2}{2}\right) \tag{S21}$$

where $\hat{y}$ is the coordinate within the thickness of the flow, $\rho$ is the fluid density, $g$ is acceleration due to gravity, $\alpha$ is the inclination angle of the surface, $\eta$ is the fluid dynamic viscosity, and $\tau$ is the thickness of the flow. The maximum flow velocity occurs at the surface where $\hat{y} = \tau$:

$$u(\hat{y})|_{\hat{y}=\tau} = \frac{\rho g}{2\eta}\tau^2 \tag{S22}$$

when the surface is vertical, $\sin\alpha = 1$. Over the duration of an exposure, where an element of photoresist traverses across the light projections, the maximum tolerable displacement of such element of photoresist due to flow should be less than the length of one voxel in the tangential direction. The conditions and parameters in this work are: $\rho = 1.1$ kg m$^{-3}$, $\tau = 3$ mm, and the maximum tolerable flow velocity is 1.1 µm s$^{-1}$, derived from the maximum displacement of the photoresist being equal to the arc length of one angular sample and from the time being equal to the duration of light exposure for the this element of photoresist from start to finish of the print. For the maximum tolerable flow velocity, it was assumed that $\delta\theta = 0.5°$ and $v_{pr} = 0.8$ mm s$^{-1}$. Solving for the minimum dynamic viscosity, where $\sin\alpha = 1$ always, $\eta \geq 43.7$ Pa s. Note that the wrapped photoresist is not always in a vertical orientation, so this estimate is likely to be conservative.

For this work, we used a gel photoresist which has near infinite viscosity. This means there is very small displacement due to flow. However, future implementations of R2R CAL could utilize liquid photoresist formulations, provided that they have sufficient viscosity to resist gravity driven flow, to expand the scope of materials beyond what is able to be thermoreversibly gelled.

### *S14. Electronic synchronization*

For each sinogram image, $M$ depth-binned images are displayed on the DMD. For each depth-binned image, the ETL must be driven to the optical power corresponding to the focus depth. This required synchronization of the drive signals for the DMD and ETL. A microcontroller was used as the source for both signals shown in Figure S12 and described as follows. The ETL was controlled with serial signals as described by the manufacturer to modulate the optical power. To begin the display, the ETL serial signal was sent. Immediately after this, a pulse was sent to the DMD to trigger it to advance by one frame in the sequence. However, on the DMD controller, a short trigger delay was added to allow for the ETL to settle at the specified optical power before displaying the image. As long as the first image in the sequence was aligned with the first optical power in the sequence of $N$ optical power settings, this scheme ensured synchronization and zero drift over time.

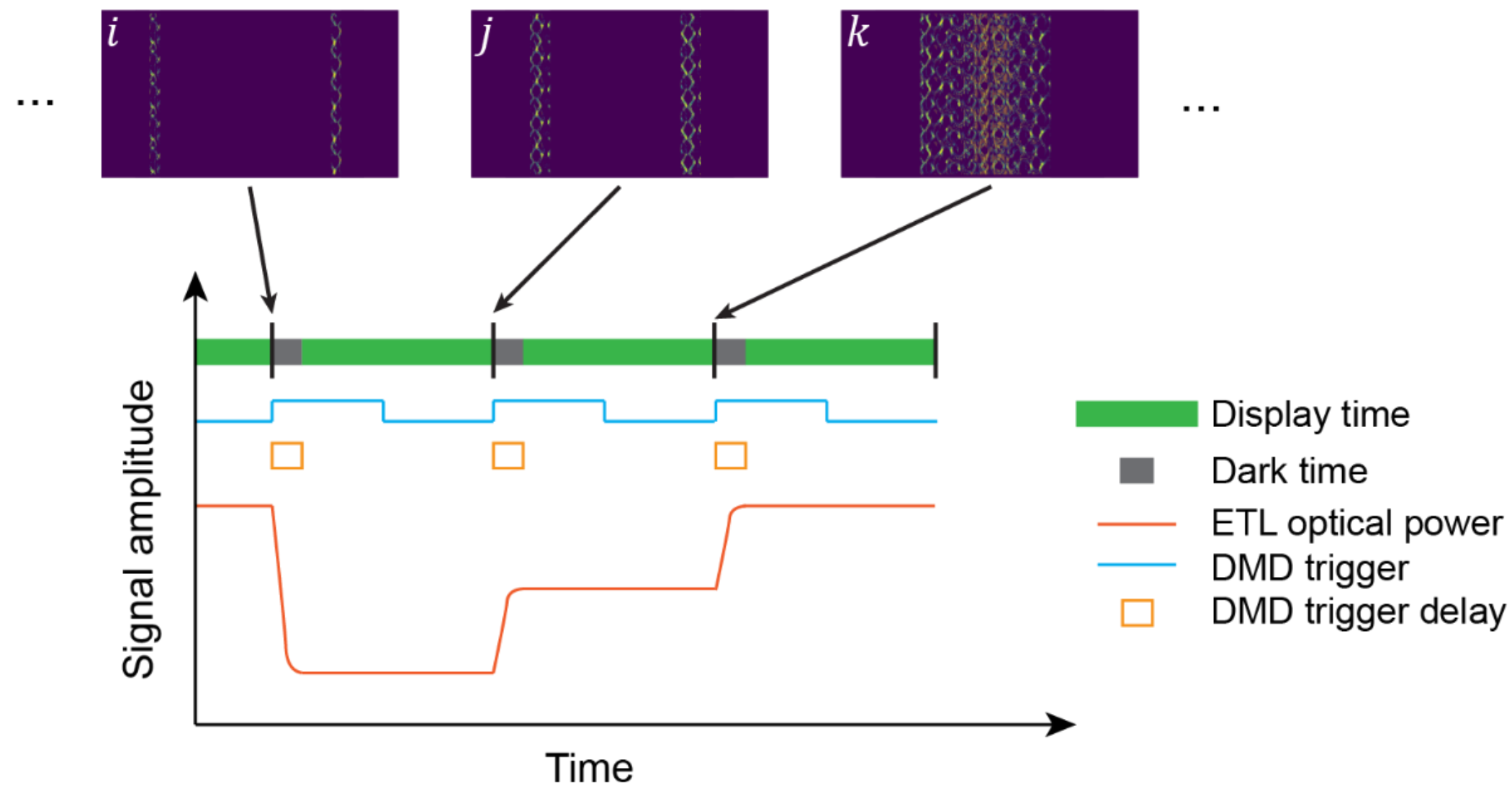

Figure S12. Schematic of ETL and DMD electronic synchronization scheme